%% file: EP241021a-rev2-nobold.tex
\documentclass{aa}

\usepackage{graphicx}
\usepackage{xcolor}
\usepackage[normalem]{ulem}
\usepackage[urlcolor =blue,colorlinks=true,linkcolor=blue,citecolor=blue]{hyperref}
\usepackage[section]{placeins}

\usepackage{siunitx}
\usepackage{booktabs}
\usepackage{longtable}
\usepackage{caption}
\usepackage[acronym]{glossaries}
\usepackage{csquotes}

\newcommand{\event}{EP241021a\xspace}
\newcommand{\mosfit}{\texttt{MOSFiT}}
\newcommand{\dynesty}{\texttt{dynesty}}
\newcommand{\default}{\texttt{default}}
\newcommand{\magni}{\texttt{magni}}
\newcommand{\magnetar}{\texttt{magnetar}}
\newcommand{\csmni}{\texttt{csmni}}
\newcommand{\csm}{\texttt{csm}}

\definecolor{blazeorange}{rgb}{1.0, 0.4, 0.0}
\definecolor{seagreen}{rgb}{0.18, 0.55, 0.34}
\definecolor{darkgreen}{rgb}{0.08, 0.45, 0.2}
\definecolor{rufous}{rgb}{0.66, 0.11, 0.03}
\definecolor{royalfuchsia}{rgb}{0.79, 0.17, 0.57}
\definecolor{scarlet}{rgb}{1.0, 0.13, 0.0}
\definecolor{royalpurple}{rgb}{0.47, 0.32, 0.66}

\usepackage{txfonts}
\begin{document}

   \title{The curious case of EP241021a: Unraveling the mystery of its exceptional rebrightening
   }

\titlerunning{The exceptional rebrightening of EP241021a}

   \author{Malte Busmann
          \inst{1}\fnmsep\thanks{Recipient of a Wübben Stiftung Wissenschaft Student Grant}
          \and
          Brendan O'Connor\inst{2}\fnmsep\thanks{McWilliams Fellow}
          \and
          Julian Sommer\inst{1}
           \and
          Daniel Gruen\inst{1,3}
           \and
          Paz Beniamini\inst{4,5,6}
           \and
          Ramandeep Gill\inst{7,5}
           \and
          Michael J. Moss\inst{8}
          \and
          Antonella Palmese\inst{2}
           \and
          Arno Riffeser\inst{1,10}
           \and
          Yu-Han Yang\inst{11}
           \and
          Eleonora Troja\inst{11}
           \and
          Simone Dichiara\inst{12}
           \and
          Roberto Ricci\inst{11,13}
           \and
          Noel Klingler\inst{8,14,15}
           \and
          Claus Gössl\inst{1}
           \and
          Lei Hu\inst{2}
           \and
          Arne Rau\inst{10}
          \and
          Christoph Ries\inst{1}
          \and
          Geoffrey Ryan\inst{16}
          \and
          Michael Schmidt\inst{1}
          \and
          Muskan Yadav\inst{11}
           \and
          Gregory R. Zeimann\inst{17}
          }

   \institute{
            University Observatory, Faculty of Physics, Ludwig-Maximilians-Universität München, Scheinerstr. 1, 81679 Munich, Germany\\
            \email{m.busmann@physik.lmu.de}
         \and
             McWilliams Center for Cosmology and Astrophysics, Department of Physics, Carnegie Mellon University, Pittsburgh, PA 15213, USA\\
             \email{boconno2@andrew.cmu.edu}
        \and Excellence Cluster ORIGINS, Boltzmannstr. 2, 85748 Garching, Germany
        \and Department of Natural Sciences, The Open University of Israel, P.O Box 808, Ra'anana 43537, Israel
        \and Astrophysics Research Center of the Open university (ARCO), The Open University of Israel, P.O Box 808, Ra'anana 43537, Israel
        \and Department of Physics, The George Washington University, Washington, DC 20052, USA
        \and Instituto de Radioastronom\'ia y Astrof\'isica, Universidad Nacional Aut\'onoma de M\'exico, Antigua Carretera a P\'atzcuaro $\#$ 8701,  Ex-Hda. San Jos\'e de la Huerta, Morelia, Michoac\'an, C.P. 58089, M\'exico
        \and Astrophysics Science Division, NASA Goddard Space Flight Center, Mail Code 661, Greenbelt, MD 20771, USA
        \and NASA Postdoctoral Program Fellow, NASA Goddard Space Flight Center, Greenbelt, MD 20771, USA
        \and Max-Planck-Institut für Extraterrestrische Physik, Giessenbachstraße 1, 85748 Garching, Germany
        \and Department of Physics, University of Rome ``Tor Vergata'', via della Ricerca Scientifica 1, I-00133 Rome, Italy
        \and Department of Astronomy and Astrophysics, The Pennsylvania State University, 525 Davey Lab, University Park, PA 16802, USA
        \and INAF-Istituto di Radioastronomia, Via Gobetti 101, I-40129 Bologna, Italy
        \and Center for Space Sciences and Technology, University of Maryland, Baltimore County, Baltimore, MD 21250, USA
        \and Center for Research and Exploration in Space Science and Technology, NASA/GSFC, Greenbelt, Maryland 20771, USA
        \and Perimeter Institute for Theoretical Physics, Waterloo, Ontario N2L 2Y5, Canada
        \and University of Texas, Hobby–Eberly Telescope, McDonald Observatory, TX 79734, USA
             }



  \abstract
   {Fast X-ray Transients (FXTs) are a rare and poorly understood phenomenon with a variety of possible progenitors. The launch of the \textit{Einstein Probe} (EP) mission has facilitated a rapid increase in the real-time discovery and follow-up of FXTs.}
   {We focus on the recent EP discovered transient EP241021a, which shows a peculiar panchromatic behavior, with the aim to understand it origin. 
   }
   {We obtained optical and near-infrared multi-band imaging and spectroscopy with the Fraunhofer Telescope at Wendelstein Observatory, the Hobby-Eberly Telescope, and the Very Large Telescope of the new EP discovered transient EP241021a over the first 100 days of its evolution. 
   }
   {EP241021a was discovered by EP as a soft X-ray trigger, but was not detected at gamma-ray frequencies. The observed soft X-ray prompt emission spectrum is consistent with non-thermal radiation, which requires at least a mildly relativistic outflow with bulk Lorentz factor $\Gamma$\,$\gtrsim$\,$4$. The optical and near-infrared lightcurve displays a two component behavior where an initially fading component $\sim$\,$t^{-1}$ turns to a rise steeper than $\sim$\,$t^{3}$ after a few days before peaking at an absolute magnitude $M_r$\,$\approx$\,$-21.8$ mag and quickly returning to the initial decay. Standard supernova models are unable to reproduce either the absolute magnitude or rapid timescale ($<$\,$2$ d) of the rebrightening. The X-ray, optical and near-infrared spectral energy distributions display a red color $r$\,$-$\,$J$\,$\approx$\,$0.8$ mag, and point to a non-thermal origin ($\sim$\,$\nu^{-1}$) for the broadband emission. By considering a gamma-ray burst as a plausible scenario, we favor a refreshed shock as the cause of the  rebrightening.
   This is consistent with the inference of an at least mildly relativistic outflow based on the prompt trigger. }
   {Our results suggest a link between EP discovered FXTs and gamma-ray bursts, despite the lack of gamma-ray detections for the majority of EP transients.}


   \maketitle
%

\section{Introduction}

Recent years have seen a variety of non-gamma-ray triggered discoveries of gamma-ray burst (GRB) afterglows \citep[e.g.,][]{Lipunov2022,HoOrphans,Perley2024,Li2024,Srinivasaragavan2025}, largely through wide-field optical surveys such as the Zwicky Transient Facility (ZTF), though other attempts have been made \citep{Freeburn2024}. These ``orphan'' afterglows \citep{Nakar2002,Huang2002,Rhoads2003} are generally observed as fast fading red transients and found at cosmological distances around $z$\,$\approx$\,$1$ \citep[e.g.,][]{HoOrphans}. A general issue, due largely to the lack of direct knowledge of the onset time of the initial GRB or afterglow, is whether these events truly lack gamma-ray emission, whether it was simply missed by gamma-ray monitors, or whether the gamma-ray emission is simply underluminous or having a low peak energy $E_\textrm{p}$. They are often suggested to be mildly relativistic outflows \citep[e.g.,][]{Perley2024,Li2024,Srinivasaragavan2025}, though slightly off-axis viewing angles cannot be ruled out as an alternative explanation \citep[e.g.,][]{Li2024}.

The \textit{Einstein probe} (EP) is a new soft X-ray mission \citep{EP2015,EP2022,Yuan2025} with wide-field capabilities.
The Wide-field X-ray Telescope (WXT) has an instantaneous field-of-view (FOV) of 3,600 deg$^2$ observing in the soft X-ray band between $0.4$\,$-$\,$4.0$ keV\@.
This revolutionary wide-field X-ray survey capability, and on-board triggering, is able to unlock the previously hidden transient X-ray sky, acting in a similar fashion for X-ray transients as gamma-ray monitors such as the \textit{Neil Gehrels Swift Observatory} \citep{Gehrels2004} and \textit{Fermi Gamma-ray Space Telescope} \citep{Meegan2009} do for GRBs.

In the first year since its launch, EP has rapidly identified a number of rare fast X-ray transients (FXTs; \citealt{Alp2020,Quirola2022,Quirola2023,Quirola2024}), allowing for key follow-up at other wavelengths. Many of these on-board EP/WXT triggers are found to be extragalactic transients. These include subthreshold events that are later identified by gamma-ray satellites in a ground-based analysis but do not trigger on-board the gamma-ray spacecraft, such as EP240219a \citep{Yin2024}, EP240315a/GRB 240315C \citep{Levan2024,Liu2024,Gillanders2024,Ricci2025}, and EP240801a \citep{Jiang2025}. However, many extragalactic EP transients are not detected by any other high-energy monitor, such as EP240408a \citep{OConnor2025,Zhang2025}, EP240414a \citep{Srivastav2024,vanDalen2024,Bright2024,Sun2024}, and EP250108a \citep{Rastinejad2025EP,Eyles-Ferris2025EP,Srinivasaragavan2025EP0108a,Li2025}. The EP/WXT triggers for some of these transients robustly constrain the peak energy $E_\textrm{p}$ of their prompt emission to soft X-rays \citep[e.g., EP240414a, EP240801a, EP250108a;][]{Liu2024,Jiang2025,Li2025}. This can help explain the lack of clear gamma-ray detections in at least a handful of sources. 
Comprehensive multi-wavelength follow-up is urgently required for revealing the nature and astrophysical diversity of these events.

Of particular interest regarding EP240414a ($z$\,$=$\,$0.4$)
are its multiple optical emission components \citep{Srivastav2024,vanDalen2024,Sun2024}, including clear spectroscopic evidence for a supernova (SN) at late times, see \citet{vanDalen2024}.
The initially fading optical lightcurve  displayed a rapid rebrightening after a few days with possible interpretations including a refreshed shock \citep{Srivastav2024}, a ``failed'' jet breakout due to an extended circumstellar envelope 
\citep{Hamidani2025}, the combination of a cocoon and supernova ejecta interactions with the surrounding circumstellar material \citep{vanDalen2024}, or an off-axis jet-cocoon system \citep{Zheng2025}. It is is also possible that EP240414a represents an entirely different class of transient that had not been previously observed.

In this manuscript, we present our multi-wavelength campaign of the recently discovered EP241021a.
We find that EP241021a presents a similar behavior to EP240414a in its multiple, clearly distinct emission episodes.
Localized to redshift $z$\,$=$\,$0.748$ \citep{VLTz,GTCz,Keckz,Shu2025}, the optical lightcurve of EP241021a was discovered as an initially fading source, but quickly exhibited a luminous rebrightening  with a fast timescale of only a few days to an absolute magnitude of $M_r$\,$\approx$\,$-21.8$ mag.
We analyze the X-ray, optical, and near-infrared dataset of EP241021a to determine whether it requires multiple emission components or can potentially be explained by the same outflow.
We further connect EP241021a to the previous peculiar EP transient EP240414a and discuss possible interpretations for their multiwavelength behavior.

Throughout the manuscript we adopt a standard $\Lambda$CDM-cosmology \citep{Planck2020} with $H_0$\,$=$\,$67.4$ km s$^{-1}$ Mpc$^{-1}$, $\Omega_\textrm{m}$\,$=$\,$0.315$, and $\Omega_\Lambda$\,$=$\,$0.685$.
We also adopt the convention for the flux density $F_\nu$\,$\propto$\,$t^\alpha\nu^\beta$ where $\alpha$ is the temporal index and $\beta$ is the spectral index.
All upper limits are reported at the $3\sigma$ level and all magnitudes are in the AB system.

\section{Observations}


\subsection{Prompt X-ray Trigger and Gamma-ray Limits}
\label{sec:trigger}

The Wide-field X-ray Telescope (WXT) onboard the \textit{Einstein Probe} (EP; \citealt{EP2015,EP2022,Yuan2025}) triggered on EP241021a on 2024-10-21 at 05:07:56 UT \citep{EP241021aWXT}.
The source had a duration of $\sim$100 s with a time-averaged X-ray flux ($0.5$\,$-$\,$4.0$\,keV) of $3.31^{+1.26}_{-0.86}\times10^{-10}$\,erg\,cm$^{-2}$~s$^{-1}$, see \citet{Shu2025} for a refined analysis and discussion of the EP/WXT trigger. 
The soft X-ray fluence is approximately $\sim$\,$3.3\times10^{-8}$\,erg\,cm$^{-2}$, yielding an estimate of the isotropic-equivalent energy ($0.5$\,$-$\,$4.0$\,keV; rest frame) of $(4.3^{+2.3}_{-1.8})\times10^{49}$\,erg at $z$\,$=$\,$0.748$ \citep{Shu2025}. In computing the K-correction $k(z)$\,$=$\,$(1+z)^{-\Gamma-2}$ to convert the observed emission to the rest-frame $0.5$\,$-$\,$4.0$\,keV bandpass, we adopted a powerlaw spectrum $N(E)$\,$\propto$\,$E^{\Gamma}$ with photon index $\Gamma$\,$=$\,$-1.80^{+0.57}_{-0.54}$ \citep{Shu2025} in the observer frame $0.5$\,$-$\,$4.0$\,keV energy range as reported for the WXT trigger \citep[][]{EP241021aWXT}. 

At the time of the EP trigger \textit{Konus-Wind} was observing the entire sky, but did not detect EP241021a in gamma-rays \citep{KonusGCN38034}. The 90\% confidence upper limit to the peak gamma-ray flux ($20$\,$-$\,$1,500$ keV) is $<$\,$2.5\times10^{-7}$ erg cm$^{-2}$ s$^{-1}$ assuming a timescale of 2.944 s \citep{KonusGCN38034}.
We adopt a typical long GRB spectrum characterized by a Band function \citep{Band1993} with peak energy $E_\textrm{p}$\,$=$\,$300$ keV and low-energy and high-energy photon indices $-1$ and $-2.5$, respectively \citep[e.g.,][]{Nava2011}. 
This bolometric correction yields a 90\% confidence upper limit to the isotropic-equivalent gamma-ray energy of $\lesssim$\,$10^{51}$ erg in the rest frame $1$\,$-$\,$10,000$ keV energy range assuming a timescale of 2.944 s for the prompt gamma-ray emission duration. If instead we assume that the gamma-ray duration is the same as the X-ray duration ($\sim$100 s), we find a less restrictive limit of $\lesssim$\,$4\times10^{52}$ erg. Past joint EP-GRB detections have shown that the soft X-ray duration can be significantly longer than the higher energy gamma-ray emission \citep[see, e.g.,][]{Yin2024,Liu2024,Yin2025ep250404a}. 
We discuss these implications further in \S\ref{sec:prompt}.

\begin{figure}
    \centering
\includegraphics[width=\columnwidth]{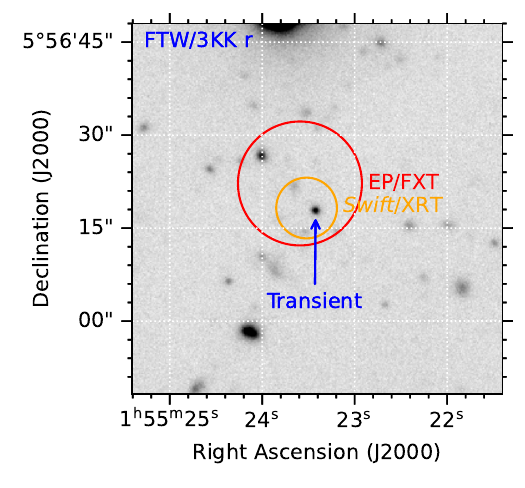}
    \caption{Finding chart of EP241021a combining all FTW exposures in $r$-band (38.1 hr total). The location of the transient (blue) lies within the X-ray localizations from \textit{Swift}/XRT (orange) and EP/FXT (red).
    }
    \label{fig:fc}
\end{figure}

\begin{figure*}
    \centering
\includegraphics[width=\textwidth]{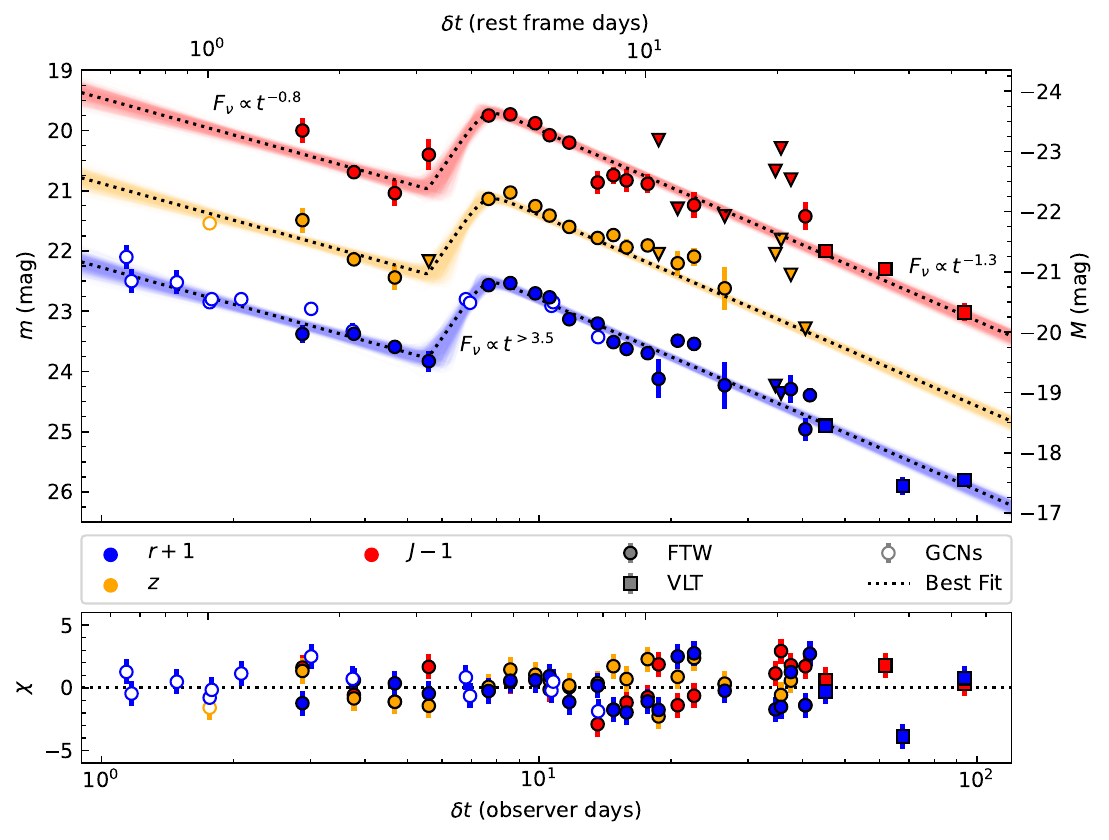}
\caption{\textbf{Top:} Optical to near-infrared lightcurve of EP241021a in the $r$, $i$, $J$ filters. We present our observations obtained with the FTW and the VLT, and supplement these data with observations reported in GCN Circulars  \citep{2024GCN.37840....1F, 2024GCN.37842....1F, 2024GCN.37844....1L, 2024GCN.37845....1R, 2024GCN.37846....1L, 2024GCN.37849....1Z, VLTz, 2024GCN.37869....1B, 2024GCN.37930....1Q, 2024GCN.37942....1F, 2024GCN.38022....1S, 2024GCN.38030....1B, 2024GCN.38071}.
The best fit of the light curve using Equation~\ref{eqn:lc} is indicated by the dotted lines and 1000 random draws from the posterior are plotted in the background. 
The Galactic extinction corrected apparent magnitude $m$ is shown on the left ordinate and the K-corrected absolute magnitude $M$ assuming $z=0.748$ on the right (see \S \ref{sec:abs}). The lower abscissa indicates the time since the EP trigger in the observer frame while the upper axis is converted to the rest frame of the source.
\textbf{Bottom:} Deviation of the data from the best fit normalized by their uncertainties. The symbols are the same as for the upper panel but with no offset applied and the x-axes are identical to the top panel.
}
    \label{fig:optlcmag}
\end{figure*}

\subsection{Fraunhofer Telescope Wendelstein (FTW)}
\label{sec:ftw}

We observed the optical and near-infrared (OIR) counterpart of \event with the Three Channel Imager (3KK; \citealt{2016SPIE.9908E..44L}) on the 2.1 m Fraunhofer Telescope at Wendelstein Observatory (FTW; \citealt{2014SPIE.9145E..2DH}) located on Mt.~Wendelstein at the northern edge of the Alps.
3KK can observe a $7\arcmin\times 7\arcmin$ FOV in three channels simultaneously.
The blue channel can observe either in the \textit{u'}, \textit{g'} or \textit{r'} band, the red channel in the \textit{i'} or \textit{z'} band, and the NIR channel supports the \textit{Y}, \textit{J}, \textit{H} and $\textit{K}_\textit{s}$ bands.
Data were obtained 22 times between 2024-10-23T02:12:17~UT (2.8 d after the EP/WXT trigger) and 2024-12-01T18:26:38~UT (43 d after the EP/WXT trigger) in the \textit{r'}, \textit{i'}, \textit{z'}, \textit{J}, and \textit{H} band.
The majority of observations were obtained simultaneously in the \textit{r'z'J} filters.

The optical CCD data are reduced and analyzed using a custom data analysis pipeline based on \citet{2002A&A...381.1095G} that applies standard image processing techniques such as bias and dark subtraction, flat-fielding, cosmic ray rejection, among other corrections. 
For the infrared CMOS images a dedicated software that is internally developed at Wendelstein Observatory is used to calculate Fowler sampled stacks from the individual sample-up-the-ramp exposure series \citep{1990ApJ...353L..33F}. This software also accounts for the CMOS non-linearity and removes fast changing bias patterns by subtracting appropriate flux conserving wavelets. These images then enter the above mentioned pipeline in the same way as bias and dark corrected CCD frames.
The images are stacked using tools from the AstrOmatic software suite \citep{1996A&AS..117..393B, 2006ASPC..351..112B, 2002ASPC..281..228B}.
To account for time dependent changes in the background, the background is modeled and subtracted for each individual frame using the tools provided by \texttt{SWarp} \citep{2002ASPC..281..228B, 1996A&AS..117..393B}.
The OIR counterpart is detected throughout our observations (see Figure~\ref{fig:fc}), which include a net 38.1 hr of exposure with the 3KK imager.
Archival observations as part of the DESI Legacy Surveys do not reveal an underlying galaxy to deep limits \citep{2024GCN.37840....1F}, and as such we do not perform difference imaging due to the lack of existing templates or a known host galaxy brightness.
Aperture photometry of the OIR counterpart was computed using \texttt{Photutils} \citep{Bradley2024} and calibrated to the 2MASS \citep{Skrutskie2006} and PS1 \citep{Chambers2016} catalogs.
We utilized an annulus placed around the circular source aperture to estimate and subtract the local background.
The photometry is tabulated in Table~\ref{tab: observationsPhot} and displayed in Figure~\ref{fig:optlcmag}.

Using the stack of all 3KK $r$-band data (Figure~\ref{fig:fc}), we localize the optical source to $\mathrm{RA, DEC\, (J2000)} = 01^\mathrm{h}55^\mathrm{m} 23^\mathrm{s}.430, +0\ang{5;56;17.86}$ with a total (systematic and statistical) positional uncertainty of $0.10\arcsec$ at $1\sigma$ CL (\SI{0.15}{\arcsec} at the $90\%$ confidence level; CL). The statistical uncertainty on the source position is $0.008\arcsec$ ($1\sigma$ CL) in both RA and DEC. The position and its statistical uncertainty is computed by fitting the point spread function of the source as a Moffat profile using the \texttt{photutils} \citep{Bradley2024} package. The systematic (absolute) astrometric tie uncertainty of sources in the field relative to Gaia EDR3 \citep{Gaia2021, 2021A&A...649A...2L, gaiaEDR3} is $0.10\arcsec$ ($1\sigma$ CL) in both RA and DEC computed using \texttt{SCAMP} \citep{2006ASPC..351..112B}. 

\subsection{Hobby-Eberly Telescope (HET)}
\label{sec:het}

\begin{figure}
    \centering
\includegraphics[width=\columnwidth]{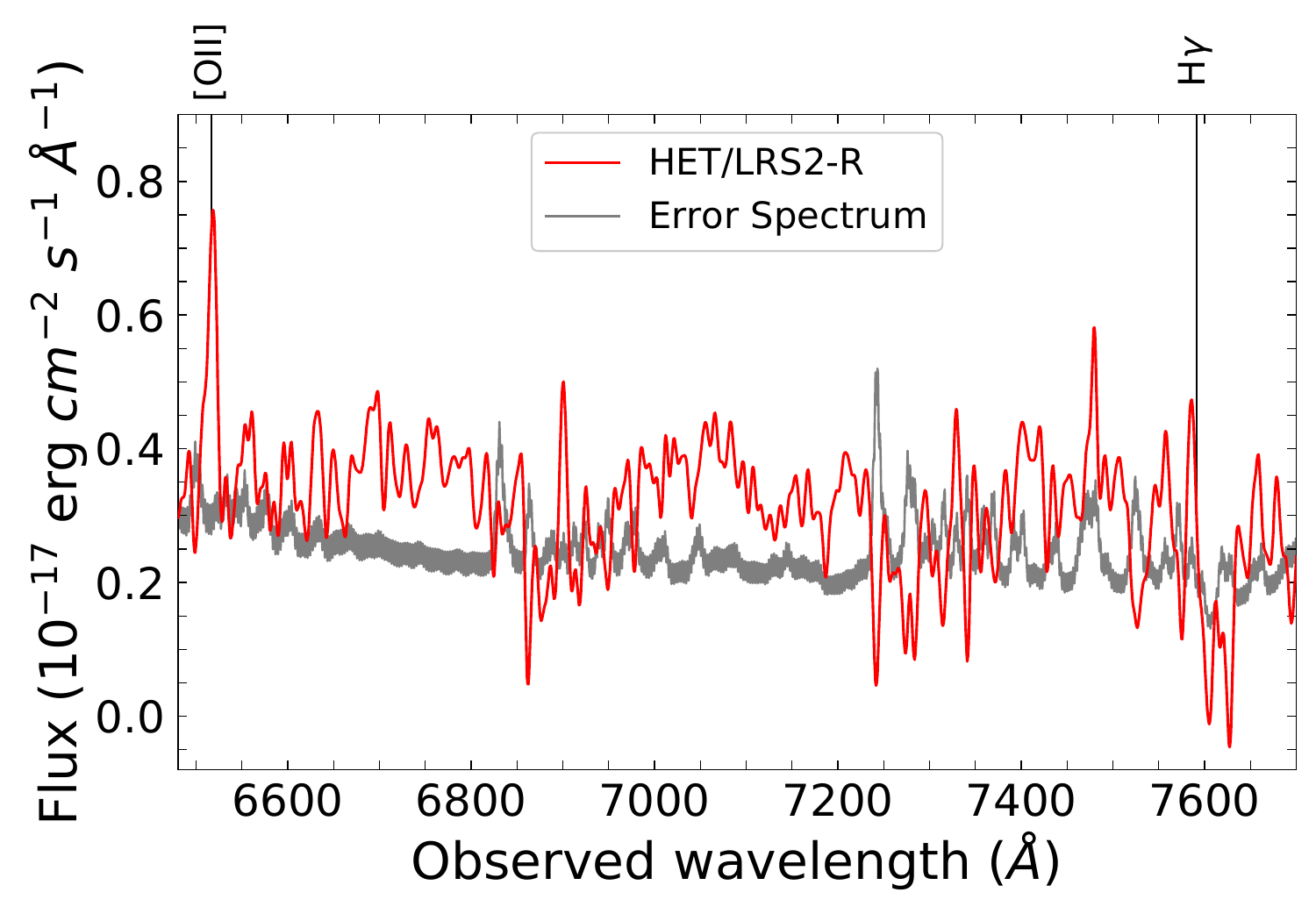}
    \caption{HET LRS2-R optical spectrum from 2024-10-29 in red and error spectrum in gray. We do not detect a clear continuum, but identify a single emission feature corresponding to [OII]$_{\lambda3729}$ at $z$\,$=$\,$0.748$. This is consistent with previous reports of emission features underlying EP241021a \citep{Shu2025}. For display purposes the data is smoothed with a Savitzky–Golay filter of 5 pixels.
    }
    \label{fig:het}
\end{figure}

We observed EP241021a with the 11 m Hobby-Eberly Telescope (HET; \citealt{1998SPIE.3352...34R, 2021AJ....162..298H}) at McDonald Observatory through program M24-3-005 (PI: Gruen) on 2024-10-25, 2024-10-29, and 2024-11-26 (see Table~\ref{tab: observationsSpec}).
The observations were scheduled using the HET's queue scheduling system \citep{2007PASP..119..556S}.
We used the  low-resolution integral-field spectrograph (LRS2; \citealt{Chonis2014,Chonis2016}) to obtain spectra in both wavelength channels with LRS2-B (370-700 nm) and LRS2-R (650-1050 nm).
LRS2-B data were obtained on 2024-10-25 and LRS2-R on both 2024-10-29 and 2024-11-26 (Table~\ref{tab: observationsSpec}).
The raw LRS2 data are initially processed with \texttt{Panacea}\footnote{\url{https://github.com/grzeimann/Panacea}}, which carries out bias subtraction, dark subtraction, fiber tracing, fiber wavelength evaluation, fiber extraction, fiber-to-fiber normalization, source detection, source extraction, and flux calibration for each channel.
The absolute flux calibration comes from default response curves and measures of the mirror illumination as well as the exposure throughput from guider images.
We extracted the flux-calibrated one-dimensional spectrum with the \texttt{LRS2Multi}\footnote{\url{https://github.com/grzeimann/LRS2Multi}} package to integrate over the $0.59\arcsec$ fibers in a $1.0$\,$-$\,$1.5\arcsec$ aperture centered on \event. 
The signal-to-noise in these observations is very low and the continuum is not clearly detected. We are only able to identify a single emission feature at 6519\,\AA, likely corresponding to [OII]$_{\lambda3729}$ at $z$\,$=$\,$0.748$, as previously reported as the redshift of underlying emission lines and absorption lines in the transient spectra \citep{VLTz,Shu2025}. Integrating over the fibers in a $1.0\arcsec$ aperture in the wavelength range centered at $6519\pm10$ \AA\, (the width of the feature in wavelength is a full-width-at-half-maximum of 5 \AA), we find a signal-to-noise of 9.3 using the LRS2-R spectrum from 2024-10-29. 
The LRS2-R spectrum from 2024-10-29, when the transient was brightest in comparison to the other epochs, is shown in Figure \ref{fig:het}. 

\subsection{Very Large Telescope (VLT)}
\label{sec:vlt}

We carried out multi-band imaging observations of EP241021a with the Very Large Telescope (VLT) at Cerro Paranal, Chile using the FORS2 \citep{Appenzeller1998} and HAWK-I \citep{Kissler-Patig2008} instruments under program 114.27LW (PI: Troja).
Data were obtained on 2024-12-05 and 2024-12-28 in $R$-band (1200 s) with FORS2, and 2024-12-05 and 2024-12-22 in $J$-band (900 s) with HAWK-I. Additional data was obtained on 2025-01-23 in both filters with integration times of 1200 s. 
The data were reduced using the standard ESO Data Reduction Pipelines and the individual images were combined using \texttt{SWarp} \citep{Bertin2010} to create a stacked image. 
Aperture photometry was performed with \texttt{SExtractor} \citep{1996A&AS..117..393B} using a circular aperture with diameter equal to twice the seeing (full width at half maximum of isolated, unsaturated point sources) in each epoch. The \texttt{SExtractor} parameters were set as described in \citet{OConnor2022}. We calibrated the photometry using isolated, unsaturated point sources in the 2MASS \citep{Skrutskie2006} and PS1 \citep{Chambers2016} catalogs for the $R$-band and $J$-band, respectively.

\subsection{\textit{Neil Gehrels Swift Observatory}}

The position of EP241021a was observed by the \textit{Neil Gehrels Swift Observatory} (hereafter \textit{Swift}; \citealt{Gehrels2004}) X-ray Telescope \citep[XRT;][]{Burrows2005} between 2024-10-24 and 2024-11-15 (ObsIDs: 21725 and 1889) for seven visits totaling 17.5 ks in Photon Counting (PC) mode.
We used the \textit{Swift}/XRT data products generator\footnote{\url{https://www.swift.ac.uk/user_objects/}} to analyze these data.
An X-ray source is detected on 2024-10-29, 2024-11-07, and 2024-11-15 and localized to RA, DEC (J2000) = $01^\mathrm{h}55^\mathrm{m} 23^\mathrm{s}.53$, $+05^\circ 56\arcmin 18.3\arcsec$ with uncertainty of $4.9\arcsec$ (90\% CL). This position is consistent with the optical localization of the transient.


The X-ray source is weak and the spectrum is not well-constrained.
A fit to the time-averaged X-ray spectrum with an absorbed powerlaw model yields a photon index of $\Gamma$\,$\approx$\,$-2.0^{+1.8}_{-0.6}$ and an unabsorbed energy conversion factor of $\approx$\,$3.3\times10^{-11}$ erg cm$^{-2}$ cts$^{-1}$ with the hydrogen column density fixed to the line-of-sight Galactic value \citep[$N_H$\,$=$\,$5\times10^{20}$ cm$^{-2}$, ][]{Willingale2013}.
The log of X-ray observations is reported in Table~\ref{tab: observationsXray}.


The \textit{Swift}  Ultra-Violet Optical Telescope (UVOT; \citealt{Roming2005}) observed EP241021a simultaneously to XRT. The source is detected only in a single epoch on 2024-10-29, as previously reported by \citet{GCN.37990}.
We extracted upper limits at the source position in all other epochs using the \texttt{uvotsource} task within the \texttt{HEASoft} software.
The results are presented in Table~\ref{tab: observationsPhot}.

\section{Analysis and Results}

\subsection{Temporal Evolution}
\label{sec:temporal}

The optical and near-infrared lightcurve of EP241021a displays two clear emission episodes (Figure~\ref{fig:optlcmag}). Around a day after the EP trigger the optical emission was found to be rapidly fading \citep[e.g.,][]{2024GCN.37840....1F, 2024GCN.37842....1F, 2024GCN.37844....1L, 2024GCN.37845....1R, 2024GCN.37846....1L}, but at $\sim$\,7 d (observer frame; e.g., \citealt{2024GCN.37930....1Q, 2024GCN.37942....1F}) the transient was observed to have rapidly brightened by $\sim$\,$1$ mag, peaking by $\sim$\,8 d (observer frame).
After the peak, the transient again returned to a steep decay.
Our observations with FTW began at 2.9 d (observer frame), see Table~\ref{tab: observationsPhot}.
We supplemented our dataset with early time photometry reported in GCN Circulars\footnote{\url{https://gcn.nasa.gov/}}, as shown in Figure~\ref{fig:optlcmag}.

As described above, the lightcurve behavior can be clearly separated into two components (Figure~\ref{fig:optlcmag}), which we refer to as the first and second component, respectively.
We separate these by defining the first component as the early fading behavior at $<$\,$6$ d (observer frame), and the second component as all other OIR data obtained between $\sim$\,$6$\,$-$\,$100$ d.
To constrain the temporal behavior, we model the OIR lightcurve, including both components, with a double smoothly broken powerlaw $F_\nu$\,$\propto$\,$t^{\alpha}$ with temporal slope $\alpha$, where $\alpha_1$ refers to the initial fading segment, $\alpha_2$ is the rising slope, and $\alpha_3$ is the post-peak decay. The full equation used in the fit is:
\begin{align}
\label{eqn:lc}
        f(t,\nu) = A\nu^{\beta}  \left( \frac{t}{t_1} \right)^{\alpha_1}\! 
        \left[ \frac{1}{2} \! \left( 1 \!+\! \frac{t}{t_1} \right)^\frac{1}{\Delta_1}\! \right]^{(\alpha_2 - \alpha_1)\Delta_1} \! 
        \left[ \frac{1}{2}\! \left( 1 \!+ \!\frac{t}{t_2} \right)^\frac{1}{\Delta_2}\! \right]^{(\alpha_3 - \alpha_2)\Delta_2},
\end{align}
where $t_1$ (rise) and $t_2$ (peak) are the temporal break times in units of observer frame days, $\Delta_1$ and $\Delta_2$ are the smoothness parameters, and $\beta$ is the spectral index. In Equation \ref{eqn:lc}, we have assumed the source's spectral energy distribution (SED) is described by a single powerlaw $\nu^{\beta}$ with no additional curvature due to added dust. We discuss the source SED further in \S \ref{sec:spectral} to justify the choice of a single powerlaw with no additional intrinsic dust, beyond correcting for Galactic extinction \citep{Schlafly2011}. 
For the fit, $\nu$ is provided in units of $10^{14}\,\si{\hertz}$.
The priors for the fits are listed in Table~\ref{tab:priors-light-curve} in Appendix \ref{Appendix Fit Results}.

We fit Equation \ref{eqn:lc} to the full lightcurve shown in Figure \ref{fig:optlcmag} using all filters ($rzJ$) simultaneously where the offset between filters is determine by the optical to near-infrared spectral index $\beta$ and the multiplicative constant $A$ (see Equation \ref{eqn:lc}). The photometry is corrected for Galactic extinction $E(B-V)=0.045$ mag \citep{Schlafly2011}. 
Parameter estimation was performed using the \texttt{emcee} \citep{emcee} and \texttt{ChainConsumer} \citep{2016JOSS....1...45H} packages.

The best fit results of our temporal fit to the lighcurves are shown in Figure~\ref{fig:optlcmag}. The initial segment shows a decay of $\alpha_1$\,$=$\,$-0.81\pm0.07$, followed by a steep rise of $\alpha_2$\,$=$\,$3.9^{+5.1}_{-0.2}$.
The exact rising phase is not well constrained by the data, but the 90\% confidence lower limit to the rising slope is found to be $\alpha_2$\,$\gtrsim$\,$3.5$. The rise starts at $5.5^{+0.4}_{-0.3}$ d and then peaks at $7.8^{+0.2}_{-0.3}$ d followed again by another decline of $\alpha_3$\,$=$\,$-1.27\pm0.03$. For the full corner plot, see Figure~\ref{fig:power-laws-corner}.
The reduced chi-squared value for this fit is $\chi^2 / \mathrm{dof} = 164 / 77 = 2.13$.

If instead we require that the initial decaying slope $\alpha_1$ and final decaying slope $\alpha_3$ are fixed to the same value, we find $\alpha_1$\,$=$\,$\alpha_3$\,$=$\,$-1.22\pm0.03$.
The fit and its residuals are shown in Figure~\ref{fig:optlcmag-same-slopes},  and the full corner plot is shown in Figure \ref{fig:power-laws-corner-same-slopes}.
This model provides a slightly worse description of the multi-band OIR data, as can be seen in the residuals in Figure~\ref{fig:optlcmag-same-slopes}. This is due to the reduced flexibility in the model. The reduced chi-squared value for this fit is also slightly worse with $\chi^2 / \mathrm{dof} = 201 / 76 = 2.58$. We therefore favor the fit shown in Figure~\ref{fig:optlcmag} where $\alpha_1$ and $\alpha_3$ are allowed to be distinct values.


\begin{figure*}
    \centering
\includegraphics[width=\textwidth]{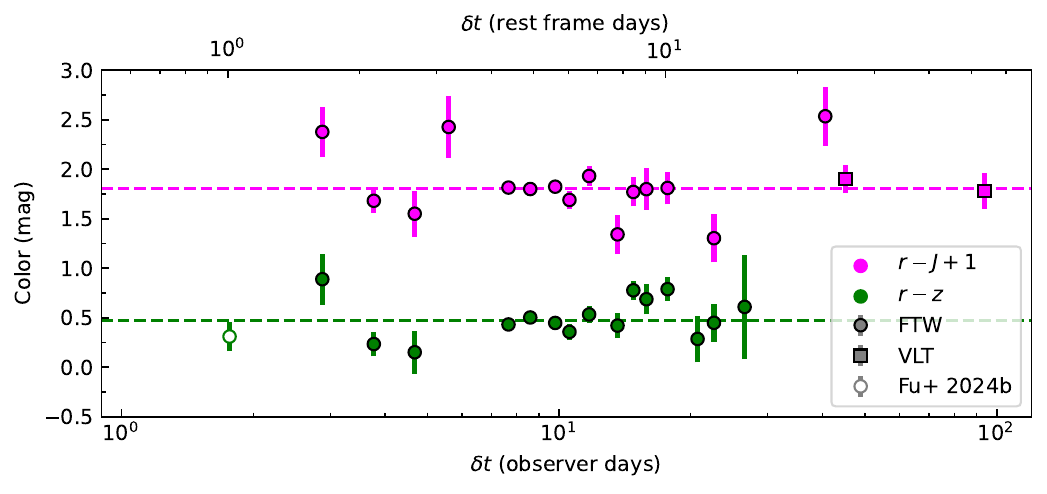}
\caption{Evolution the $r$\,$-$\,$z$ and $r$\,$-$\,$J$ colors of EP241021a. The earliest color information was provided in the GCN circular by \citet{2024GCN.37840....1F}. All other datapoints showing the color evolution come from our FTW observations (\S \ref{sec:ftw}).
}
    \label{fig:colors}
\end{figure*}

We note that the $R$-band photometry obtained with the VLT at 68 and 94 d (\S\ref{sec:vlt}) after the EP trigger appear consistent with a flat evolution. We caution that this may be due to variable seeing impacting the photometry at 68 d. In particular, the $R$-band measurement at 68 d was obtained under poor seeing (full width at half maximum of $\sim$\,$1.7\arcsec$), whereas all other VLT data was obtained under better conditions with seeing of $0.5$\,$-$\,$0.9\arcsec$. An added uncertainty is the lack of knowledge of the underlying host galaxy contribution. An underlying host galaxy at $z$\,$=$\,$0.748$ is required by the emission lines identified through long-slit optical spectroscopy \citep[see][]{Shu2025}. The underlying host galaxy contribution to the measured flux would lead us to underestimate the steepness of the transient's flux decay rate. However, a longer baseline is required to conclusively determine the host galaxy contribution (which has an unknown brightness) and subtract it from the data.

If we assume that the flat evolution of the VLT $R$-band photometry between 68 to 94 d is caused by the underlying host galaxy (i.e., no longer probing the transient behavior) then the late-time slope $\alpha_3$ would be steeper. We therefore also performed the same temporal fit outlined above after artificially subtracting the inverse variance weighted mean of the last two data points from the lightcurve, and excluding them from the fit. This fit was only performed in the $r$ band because it is the only band where an apparent flattening is visible.
$\alpha_1$ and $\alpha_2$ are not significantly changed by this subtraction, but it does lead to a steeper decaying slope of the second segment with $\alpha_3$\,$=$\,$-1.49^{+0.07}_{-0.09}$.
A late-time template image is required to properly subtract the host contribution and confirm any additional temporal breaks in the lightcurve, which will aid in the modeling and interpretation of the source.

We likewise model the \textit{Swift}/XRT X-ray lightcurve (Table \ref{tab: observationsXray}) with a single powerlaw, which yields a temporal index $\alpha_X$\,$=$\,$-0.5^{+0.4}_{-0.2}$. This is consistent with the shallow X-ray decay ($\alpha_X$\,$=$\,$-0.31^{+0.17}_{-0.13}$) at early times measured by \citet{Shu2025} using the full EP/FXT lightcurve.
The measured X-ray decay (see also Figure~\ref{fig:gamma}) is consistent with the temporal indices ($\alpha_X$\,$\approx$\,$-0.7$ to $0$; e.g., \citealt[][]{Nousek2006,Bernardini2011}) of GRB X-ray plateaus, which are generally shallower than the expectations of a forward shock afterglow \citep{Granot2002}. 
We present further discussion of the possibility that EP241021a has an X-ray plateau in \S \ref{sec:x-rayplateau}, as also discussed in \citet{Shu2025}.

The \textit{Swift}/XRT clearly shows a shallow decay that is further supported by the refined EP/FXT lightcurve reported by \citet{Gianfagna2025,Shu2025} between $1$\,$-$\,$100$ d. The EP/FXT lightcurve shows that an X-ray rebrightening occurs simultaneously to the OIR bump \citep[e.g.,][]{Shu2025}. The refined EP/FXT X-ray lightcurve also shows a shallow plateau-like decay before the rebrightening, before steepening at later times (see \citealt{Shu2025} for further discussion). 
In any case, this motivates deep, higher cadence X-ray observations of future EP transients to better constrain their full multi-wavelength behavior.

\subsection{Spectral Evolution}
\label{sec:spectral}

In Figure~\ref{fig:colors}, we show that EP241021a displays a consistent, red color throughout its entire evolution (see also Figure~\ref{fig:optlcmag}). The photometry has been corrected for Galactic extinction $E(B-V)=0.045$ mag \citep{Schlafly2011}. We fit the $r$\,$-$\,$z$ and $r$\,$-$\,$J$ color as a function of time with a constant value for either color (as shown in Figure~\ref{fig:colors}), deriving $r$\,$-$\,$z$\,$=$\,$0.48\pm 0.03$ and $r$\,$-$\,$J$\,$=$\,$0.81\pm 0.03$ mag  with a $\chi^2$/dof\,$=$\,$2.20$ for $15$ dof and $\chi^2$/dof\,$=$\,$1.92$ for $16$ dof, respectively. Upper limits were not included in the fit because they do not constrain the color due to their large uncertainties. There is a marginal deviation from this value in a small number of datapoints (Figure ~\ref{fig:colors}), but we find this is due to the large photometric errors at these phases that are caused by low signal-to-noise detections.

\begin{figure}
    \centering
\includegraphics[width=\columnwidth]{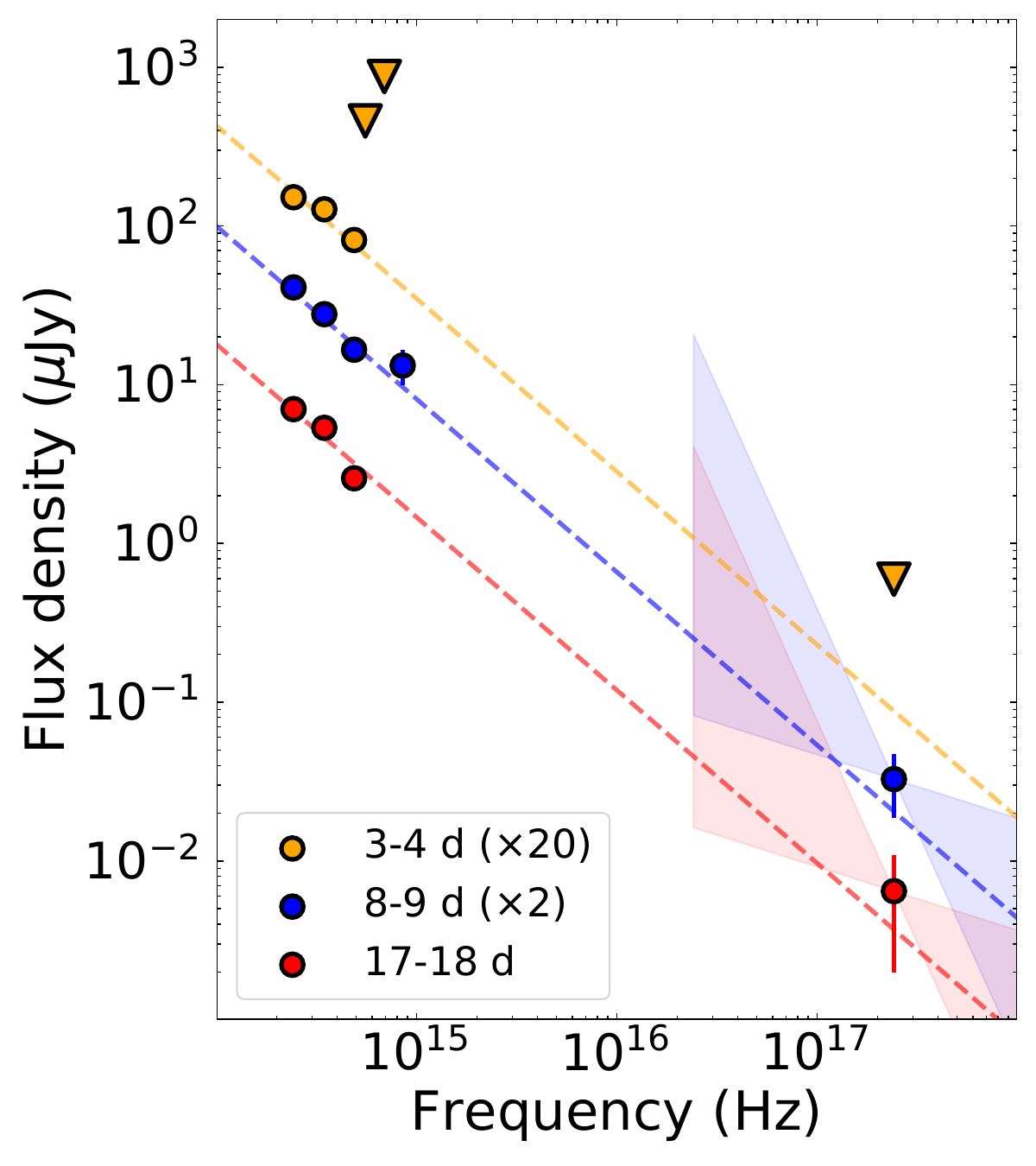}
    \caption{Multi-epoch spectral energy distribution (SED) of EP241021a from X-ray to near-infrared frequencies using \textit{Swift}/XRT and FTW. The three epochs represent pre-peak, peak, and post-peak SEDs. The uncertainty on the X-ray spectral index $\nu^{\beta}$ is shown as a shaded region with the median represented by a dotted line. The dashed lines connecting the XOIR data use the best fit OIR spectral index $\beta_\mathrm{OIR}$\,$=$\,$-1.09$. The X-ray data is consistent with extrapolating the OIR spectral index.
    }
    \label{fig:SEDs}
\end{figure}

The full OIR dataset was modeled using Equation \ref{eqn:lc} and the methods outlined in \S \ref{sec:temporal}. In Equation \ref{eqn:lc}, we adopt a single powerlaw of $\nu^\beta$ for the source's spectral energy distribution. The photometry was for Galactic extinction $E(B-V)=0.045$ mag \citep{Schlafly2011} prior to any fitting. We do not include additional dust intrinsic to the host galaxy. 
Using this fitting method, we find a OIR spectral index of $\beta_\mathrm{OIR}$\,$=$\,$-1.09^{+0.05}_{-0.06}$ (dashed lines in Figure~\ref{fig:SEDs}). The posterior for the full fit, including the spectral index, is shown in Figure \ref{fig:power-laws-corner}. 
This provides a consistent description of the optical color (Figure~\ref{fig:colors}) throughout the evolution of the transient.


We find that extrapolating the OIR spectral index to X-ray frequencies is consistent with the near-simultaneous X-ray data from \textit{Swift}/XRT. In Figure~\ref{fig:SEDs} we show the near-simultaneous broadband (X-ray to near-infrared) spectral energy distribution across multiple epochs (observer frame times of $\sim$\,$3$, $8$, and $17$ d). For the purposes of Figure~\ref{fig:SEDs}, the OIR data was shifted to the mid-time of the X-ray observations (Table~\ref{tab: observationsXray}) using the best fit OIR temporal fit (Equation \ref{eqn:lc}) as shown in Figure \ref{fig:optlcmag} (see \S \ref{sec:temporal} for details). While the \textit{Swift}/XRT data have large error bars and a poorly constrained X-ray photon index ($\Gamma$\,$=$\,$\beta-1$\,$\approx$\,$-2.0^{+1.8}_{-0.6}$),  we note that an analysis of the full EP/FXT dataset between 1.5 and 8.2 days by \citet{Gianfagna2025} yields a refined X-ray photon index of $\Gamma$\,$=$\,$1.92\pm0.22$ which is consistent with our inferred OIR spectral index.



\subsection{Absolute Magnitude}
\label{sec:abs}

In Figure~\ref{fig:optlcmag} (right side y-axis), we also demonstrate the rest-frame absolute magnitude of the OIR filters assuming a redshift of $z$\,$=$\,$0.748$. After correcting for Galactic extinction $E(B-V)=0.045$ mag, we convert the observed apparent magnitude in each filter to its rest frame equivalent bandpass (i.e., observer frame $r$-band to rest frame $r$-band). 
Assuming a powerlaw spectrum $F_\nu$\,$\propto$\,$\nu^\beta$ (\S \ref{sec:spectral}), the K-correction is given by $F_\nu$\,$\propto$\,$(1+z)^{-\beta-1}$ which for $\beta$\,$=$\,$-1.08$ is a minor correction to the absolute magnitude $k(z)$\,$=$\,$-2.5(-\beta-1)\log_{10}(1+z)$. 
We note that in the case of the observer frame $r$-band it could also be converted to the rest frame $u$-band for $z$\,$=$\,$0.748$. Using this conversion, the rebrightening episode of EP241021a peaks at a rest frame absolute magnitude of $M_r$\,$=$\,$-21.80\pm 0.11$ mag in $r$-band measured at 8.6 d in the observer frame (4.9 d in the rest frame). We have accounted for errors on the inferred spectral index (\S \ref{sec:spectral}). As the transient is found to be initially fading in the earliest observations (Figure~\ref{fig:optlcmag}), we note that it is possible that the onset of the earlier emission episode is significantly more luminous.

\section{Discussion}

\subsection{The multi-wavelength properties of EP241021a}

Here we briefly summarize the results of our analysis in terms of the multi-wavelength properties of EP241021a:
\begin{itemize}
    \item The prompt emission at soft X-ray wavelengths is consistent with the energetics of low-luminosity GRBs, while it lacked any gamma-ray detection (see \S\ref{sec:trigger} and Figure~\ref{fig:gamma}). 
    \item The OIR lightcurve rebrightens starting sometime after $\sim$\,3.2 d (rest frame) and peaks at $\sim$\,4.5 d (rest frame), see \S\ref{sec:temporal} and Figure~\ref{fig:optlcmag}. 
    \item The source has a consistently red color ($r$\,$-$\,$J$\,$=$\,$0.81\pm0.03$ mag; \S\ref{sec:spectral}) that is indicative of non-thermal emission (see Figures~\ref{fig:colors} and \ref{fig:SEDs}). 
    \item The presence of luminous X-ray and radio emission strengthens the evidence for a significant non-thermal component \citep[e.g.,][]{Yadav2025,Gianfagna2025,Shu2025}.
\end{itemize}

The main takeaway from this multi-wavelength behavior is that any scenario to explain EP241021a requires either multiple emission components, potentially due to multiple outflows, or a mechanism to produce a rebrightening phase (e.g., energy injection, long-lived central engine activity, refreshed shock, reverse shock; see \S\ref{sec:interp} for further discussion).
In the next section, we explore the possible relation to EP240414a, which has shown a similar multi-wavelength behavior.

\subsection{A new class of transients: Comparison to EP240414a}

\begin{figure*}
    \centering
\includegraphics[width=\columnwidth]{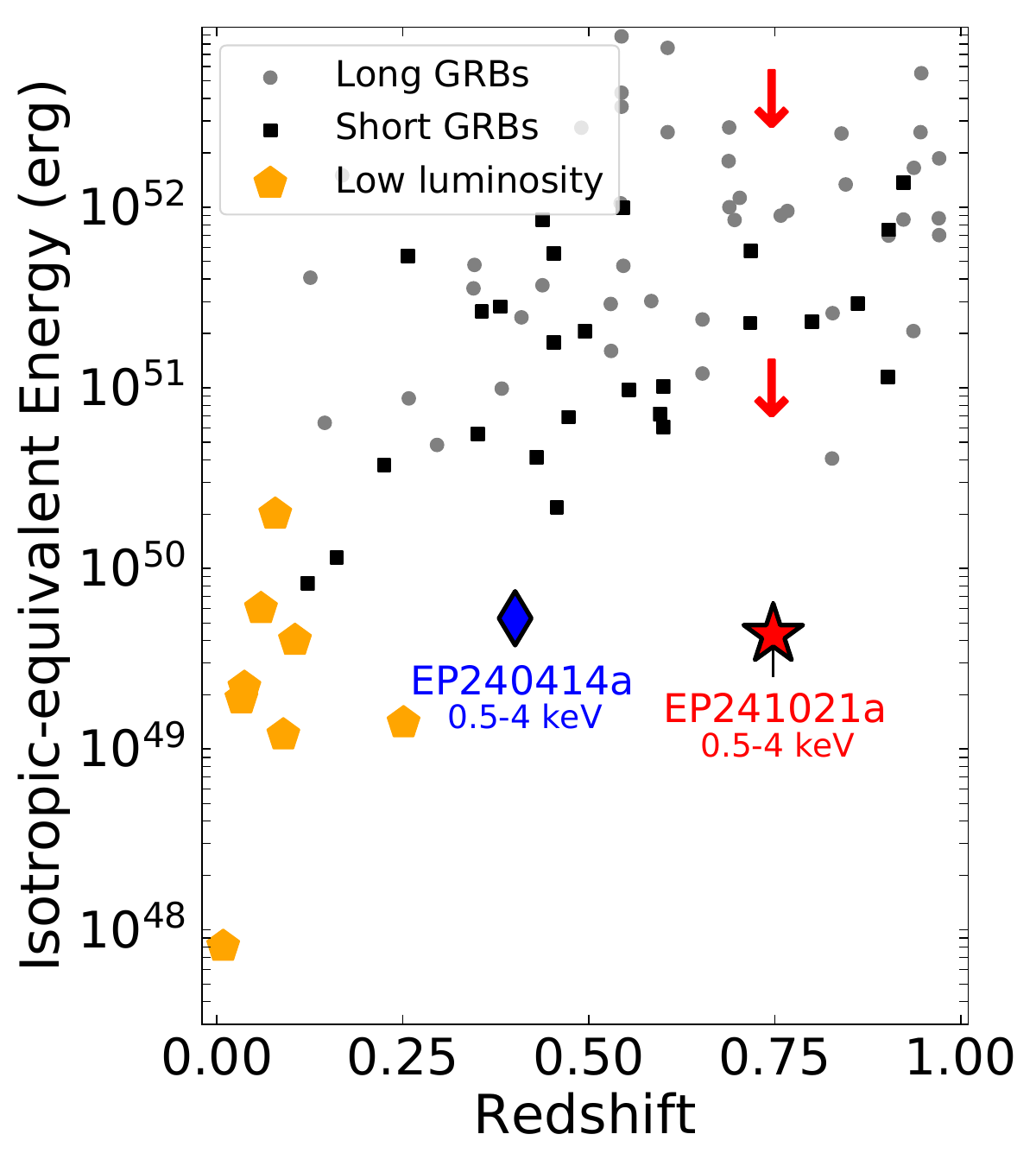}
\includegraphics[width=\columnwidth]{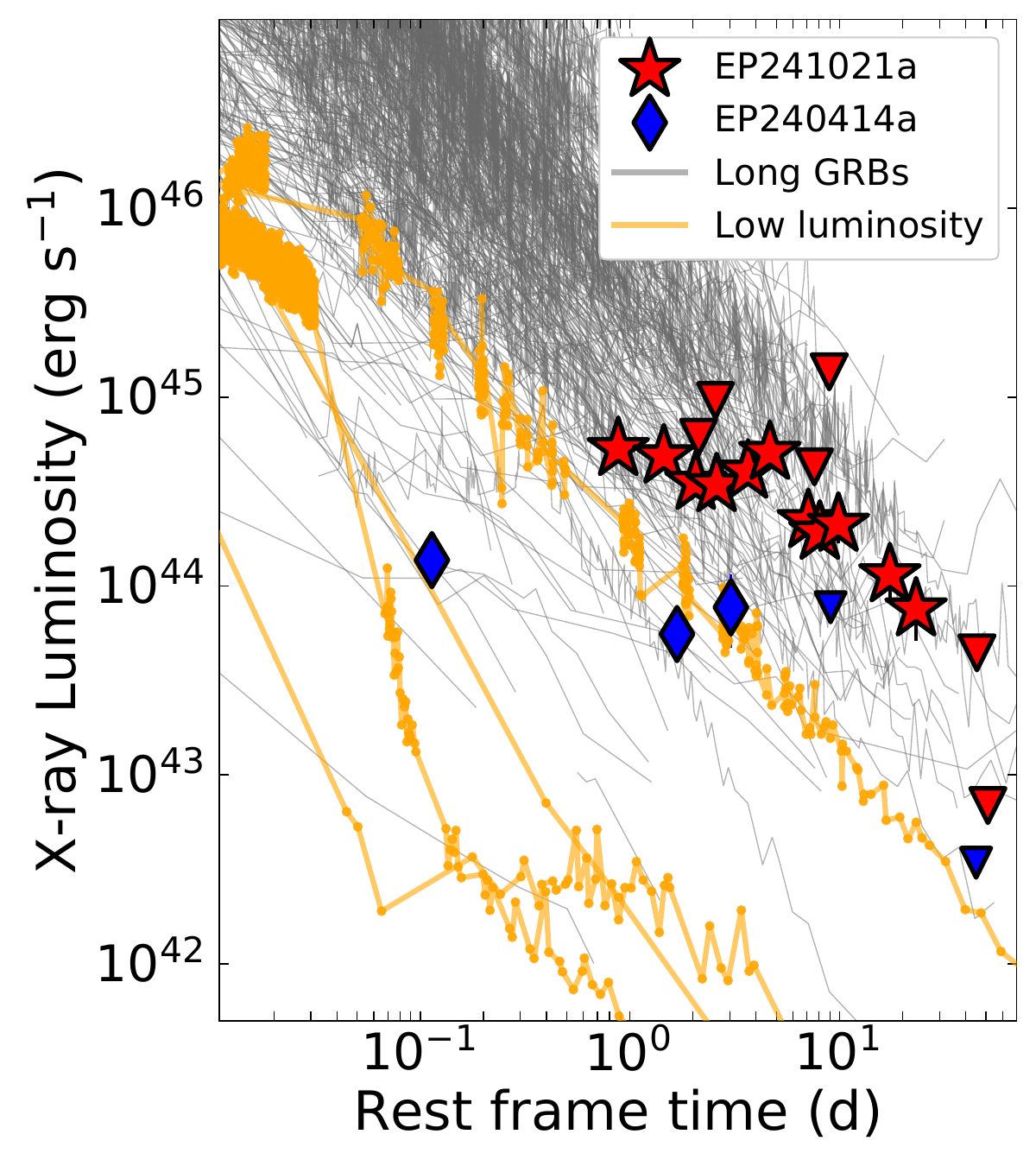}
    \caption{\textbf{Left:} Isotropic-equivalent gamma-ray energies ($1$\,$-$\,$10,000$ keV; rest frame) of both short (black squares) and long duration (gray circles) GRBs versus redshift \citep{Sakamoto2011,Lien2016,Atteia2017,OConnor2024}. A population of low-luminosity GRBs is shown by orange pentagons \citep[e.g.,][]{Iwamoto1998,Galama1998,Galama1999,Malesani2004,Sakamoto2004,Soderberg2004-020903,Soderberg2004-031203,Soderberg2006grb060218,Campana060218,Ofek2007,Starling2011,Cano2011,DElia2018,Izzo2019,Hess2021,Dichiara2022}.
    The approximate energy release ($0.5$\,$-$\,$4$ keV; rest frame) of the initial EP detection of EP241021a is shown as a red star \citep[\S\ref{sec:trigger};][]{EP241021aWXT}. We note that no bolometric correction has been applied due to the unknown spectral shape at $>$\,$4$ keV, and the presented energy is a lower bound. The downward red arrows show the gamma-ray limits for EP241021a from \textit{Konus-Wind} for gamma-ray emission durations between 3 s (bottom arrow) and 100 s (top arrow). The $1$\,$-$\,$10,000$ keV energy release of EP241021a is bounded by these values. 
    For comparison, EP240414a is also shown as a blue diamond \citep[$0.5$\,$-$\,$4$ keV;][]{Sun2024}. 
    \textbf{Right:} Rest frame X-ray lightcurves ($0.3$\,$-$\,$10$ keV) of \textit{Swift} long GRBs (gray; duration $>$\,$2$ s) and low-luminosity GRBs (orange; GRBs 060218, 100316D, 171205A, 190829A) compared to the X-ray lightcurves of EP241021a \citep[taken from][]{Shu2025} and EP240414a \citep[taken from][]{Sun2024,vanDalen2024}. 
    }
    \label{fig:gamma}
\end{figure*}

Here we summarize the properties of EP240414a ($z$\,$=$\,$0.401$), a twin to EP241021a, based on the analyses reported in the literature \citep{Srivastav2024,vanDalen2024,Bright2024,Sun2024}:
\begin{itemize}
    \item The initial EP trigger is consistent with the energetics of low-luminosity GRBs (Figure~\ref{fig:gamma}), and the peak energy $E_\textrm{p}$ is strongly constrained to the soft X-ray band \citep{Sun2024}. There was no gamma-ray detection of the prompt emission. 
    \item A redshift of $z$\,$=$\,$0.401$ was adopted based on the association to a massive host galaxy at a large projected physical offset of $\sim$26 kpc \citep{vanDalen2024,Srivastav2024}.
    \item The initially decaying OIR emission significantly rebrightened to $M_r$\,$\approx$\,$-21$ mag after 2 d (rest frame). During this phase the source displayed a red color ($r$\,$-$\,$z$\,$=$\,$0.4\pm0.1$ mag; \citealt{Srivastav2024}). A third optical emission component at late times was revealed to be a Type Ic-BL supernova \citep{vanDalen2024}. 
    \item A luminous radio counterpart revealed the presence of an at least mildly relativistic outflow with bulk Lorentz factor $\Gamma$\,$>$\,$1.6$ \citep{Bright2024}.
\end{itemize}

The exact nature of EP240414a is debated in the literature with multiple possible interpretations presented, including an afterglow origin \citep[possibly a refreshed shock;][]{Srivastav2024,Sun2024} or the combination of a cocoon and the interaction between supernova ejecta and circumstellar material  \citep{vanDalen2024}. 
\citet{Zheng2025} also argued for the jet-cocoon interaction, 
but suggested the delayed rebrightening is due to an off-axis jet. 
Instead, \citet{Hamidani2025} suggested that extended circumstellar material led to an almost failed breakout of the relativistic jet, leading it to be weakened and delayed. 


As the lack of gamma-ray emission, multiple optical emission components, long-lived and rather flat X-ray lightcurve, and luminous, late-peaking radio emission are all shared between EP240414a and EP241021a, we consider it likely these two events have a similar nature. EP240414a is conclusively associated to the death of a massive star through the presence of a Type Ic-BL supernova \citep{vanDalen2024}, and, therefore, we only consider a massive star progenitor for EP241021a and do not explore other possible explanations (e.g., a relativistic jetted tidal disruption event). We note that over the same timescale (i.e., covering the peak of a Type Ic-BL GRB-SN; \citealt{Iwamoto1998,Galama1998}), EP241021a is significantly more luminous by nearly 2 magnitudes than a Type Ic-BL supernova, which likely precluded the identification of a supernova component (as was the case for GRB 221009A and other similarly luminous afterglows; e.g., \citealt{Srinivasaragavan2023}).  
Additional EP transients have subsequently been linked to massive star explosions through the identification of Type Ic-BL supernovae, e.g., EP250108a \citep{Rastinejad2025EP,Eyles-Ferris2025EP,Srinivasaragavan2025EP0108a,Li2025} and EP250304a \citep{EP250304a-SN-GCN}.

\subsection{The Prompt Emission}
\label{sec:prompt}

The high energy detection of EP241021a and the non-thermal X-ray to near-infrared SED are reminiscent of cosmological gamma-ray bursts, which are the most commonly observed class of high-energy transients.
As such, we compare the high energy properties of EP241021a to typical gamma-ray bursts.
In Figure~\ref{fig:gamma} (left panel), we show the approximate (rest-frame) isotropic-equivalent $0.5$\,$-$\,$4$ keV energy released by EP241021a during the initial EP trigger (\S\ref{sec:trigger}; \citealt{EP241021aWXT}) versus different classes of gamma-ray bursts in the (rest frame) $1$\,$-$\,$10,000$ keV band. A bolometric correction is required for EP241021a to place it in the same energy band as the gamma-ray bursts (Figure~\ref{fig:gamma}). However, the spectral shape of EP241021a's prompt emission is unknown beyond $4$ keV due to the lack of gamma-ray detections. We note that the non-detection of EP241021a by \textit{Konus-Wind} in the $20$\,$-$\,$1,500$ keV band (\S\ref{sec:trigger}) is easily explained if the peak energy $E_p$ is low, as was the case for EP240414a \citep{Sun2024}. This is supported by the lack of gamma-ray detection by \textit{Fermi}/GBM in the $10$\,$-$\,$1,000$ keV band, see Figure 2 of \citet{Yadav2025}. In comparing the reported photon indices based on powerlaw spectral fits to the EP/WXT triggers \citep{EP241021aWXT,Sun2024,Shu2025}, the spectrum of EP241021a ($\Gamma$\,$=$\,$-1.80^{+0.57}_{-0.54}$; \citealt{Shu2025}) appears substantially harder than EP240414a ($\Gamma$\,$=$\,$-3.1\pm0.8$ at $1\sigma$ CL; \citealt{Sun2024}) which had an extremely low peak energy $E_p$\,$<$\,$1.3$ keV \citep{Sun2024}, an outlier for a gamma-ray burst. 

Instead, EP241021a does not require such a low peak energy to be consistent with the non-detection of prompt gamma-ray emission \citep[see also][]{OConnor2025}. 
Figure~\ref{fig:gamma} shows the range of isotropic-equivalent energy constraints from \textit{Konus-Wind} ($\lesssim$\,$10^{51}$\,$-$\,$4\times10^{52}$ erg; $20$\,$-$\,$1,500$ keV). Adopting the upper bound of the energy range allowed by \textit{Konus-Wind} does not provide a strong constraint on the peak energy. However, we determine that for a typical long GRB spectrum (a Band function with low-energy photon index $-1$ and high-energy photon index $-2.5$) extrapolating the prompt EP trigger ($0.5$\,$-$\,$4$ keV) with a standard bolometric correction requires $E_\textrm{p}$\,$\lesssim$\,$100$ keV to not overproduce the lower bound (see \S \ref{sec:trigger} for details) of the \textit{Konus-Wind} limits ($1$\,$-$\,$10,000$ keV) for the same spectral shape.

This is a reasonable value for the peak energy and consistent with the values observed for many long GRBs and their location on the Amati relation \citep[e.g.,][]{Amati2002,Amati2006}. The Amati relation is a correlation between the rest frame isotropic-equivalent gamma-ray energy $E_\gamma$ in the $1$\,$-$\,$10,000$ keV bandpass and the rest frame spectral peak energy $E_\textrm{p}$\,$=$\,$E_\textrm{p,obs}(1+z)$, where $E_\textrm{p,obs}$ is the observed peak energy. Using the best fit Amati relation from \citet{Amati2006}, which is given by $E_\textrm{p}$\,$=$\,$95(E_\gamma/10^{52}\,\textrm{erg})^{0.49}$ keV, we estimate a limit to the rest frame peak energy of $E_\textrm{p}$\,$\lesssim$\,$30$ keV and $\lesssim$\,$190$ keV assuming either $E_\gamma$\,$\lesssim$\,$10^{51}$ erg or $\lesssim$\,$4\times10^{52}$ erg depending on the assumed prompt emission duration between 3 s and 100 s (see \S \ref{sec:trigger} for details). This corresponds to an observed peak energy in the range $\lesssim$\,$20$ to $110$ keV. 
However, due to the limited available high-energy data, we cannot determine precisely whether the lack of gamma-rays is due to the limited sensitivity of the available gamma-ray telescopes, a low peak energy $E_\textrm{p}$, or a low gamma-ray efficiency, potentially caused by a low outflow Lorentz factor or slightly off-axis viewing angle \citep[see, e.g.,][for a discussion]{BeniaminiNakar2019,OConnor2024}.

We further note that the prompt emission spectrum (with photon index $\Gamma$\,$=$\,$-1.80^{+0.57}_{-0.54}$; \citealt{Shu2025}) of EP241021a is consistent with the expectations for non-thermal synchrotron radiation produced by gamma-ray bursts \citep[e.g.,][]{Rees1992,Preece1998,Preece2002}. The inferred photon index is within the range of observed low-energy photon indices observed from cosmological gamma-ray bursts. For example, a detailed analysis of the prompt emission spectral properties of 438 \textit{Fermi} GRBs \citet{Nava2011} identified a range of low-energy photon indices between $-2$ to $0$, and based on the Third \textit{Swift} BAT Catalog \citet{Lien2016} found a range between $-2.5$ to $-0.5$. This range of values is also found from a comprehensive spectral analysis of BATSE GRBs \citep[e.g.,][]{Preece2000,Preece2002}. These values are also typical of theoretical predictions of the synchrotron shock model \citep[e.g.,][]{Rees1992,Preece1998,Preece2002}, and deviate significantly from the expected exponential tail of thermal emission. The fundamental prediction of the synchrotron shock model is a powerlaw spectrum with low-energy photon index $dN/dE$\,$\propto$\,$E^{-3/2}$ to $E^{-2/3}$ \citep[e.g.,][]{Preece1998}, where $E^{-2/3}$ is the synchrotron ``line of death''. The inferred prompt emission spectrum of EP241021a lies within this range (within 1-sigma). We note that steeper values are observed for the high-energy photon index ranging from $-3$ to $-2$ for energies above $E_\textrm{p}$ \citep[e.g.,][]{Nava2011}. Therefore, the photon index of EP241021a is in agreement with either the low- or high-energy photon indices observed from gamma-ray bursts, and significantly differs from the predictions for thermal emission (either a rising spectral slope below the peak or the exponential decay of the Wean tail; see, e.g., \citealt{Rybicki}).

The general energetics of these two events (EP240414a and EP241021a; \citealt{Sun2024,Shu2025}) clearly lie within those of the class of low-luminosity GRBs \citep[e.g.,][]{Iwamoto1998,Galama1998,Galama1999,Malesani2004,Sakamoto2004,Soderberg2004-020903,Soderberg2004-031203,Soderberg2006grb060218,Campana060218,Ofek2007,Starling2011,Cano2011,DElia2018,Izzo2019,Hess2021,Dichiara2022,Irwin2024a}, which could be easily missed by \textit{Konus-Wind} depending on the uncertain spectral shape of the prompt emission. 
On the other hand, the X-ray emission is more luminous and long-lived than observed in standard low-luminosity GRBs (Figure~\ref{fig:gamma}; right panel), e.g., GRB 171205A \citep{DElia2018}. Moreover, the optical lightcurve is significantly different from the typically supernova dominated lightcurves of low-luminosity GRBs \citep[e.g.,][]{Soderberg2006grb060218,Campana060218,Starling2011,Cano2011,DElia2018,Izzo2019}.

\subsection{The X-ray Lightcurve}
\label{sec:x-rayplateau}

The long-lasting, shallow X-ray emission of both EP240414a \citep[e.g.,][]{Sun2024,Zheng2025} and EP241021a \citep[Figure \ref{fig:gamma};][]{Yadav2025,Shu2025,Gianfagna2025} are reminiscent of the X-ray plateaus observed from gamma-ray bursts \citep[e.g.,][]{Zhang2006,Troja2007,Dainotti2008,Rowlinson2010}. Here, we define an X-ray plateau as a decay shallower than $t^{-3/4}$ \citep[e.g.,][]{Nousek2006}, which is difficult to reproduce in the relativistic standard fireball model of GRB afterglows \citep{SariPiranHalpern1999,Wijers1999,Granot2002}. While the precise mechanism for producing X-ray plateaus is debated, multiple possible interpretations exist. In general, the preferred models usually involved energy injection from a long-lived central engine \citep{Zhang2006,Liang2006,Troja2007,Lyons2010,Rowlinson2010,Rowlinson2013}, though other interpretations exist \citep{Shen2012,Beniamini2017,Beniamini2019plateau,Dereli-Begue2022}. GRB X-ray plateaus are often found to follow correlations between their luminosity and the end time of the plateau \citep[see, e.g.,][]{Dainotti2008,Tang2019,Xu2021}.

\begin{figure}
    \centering
    \includegraphics[width=\columnwidth]{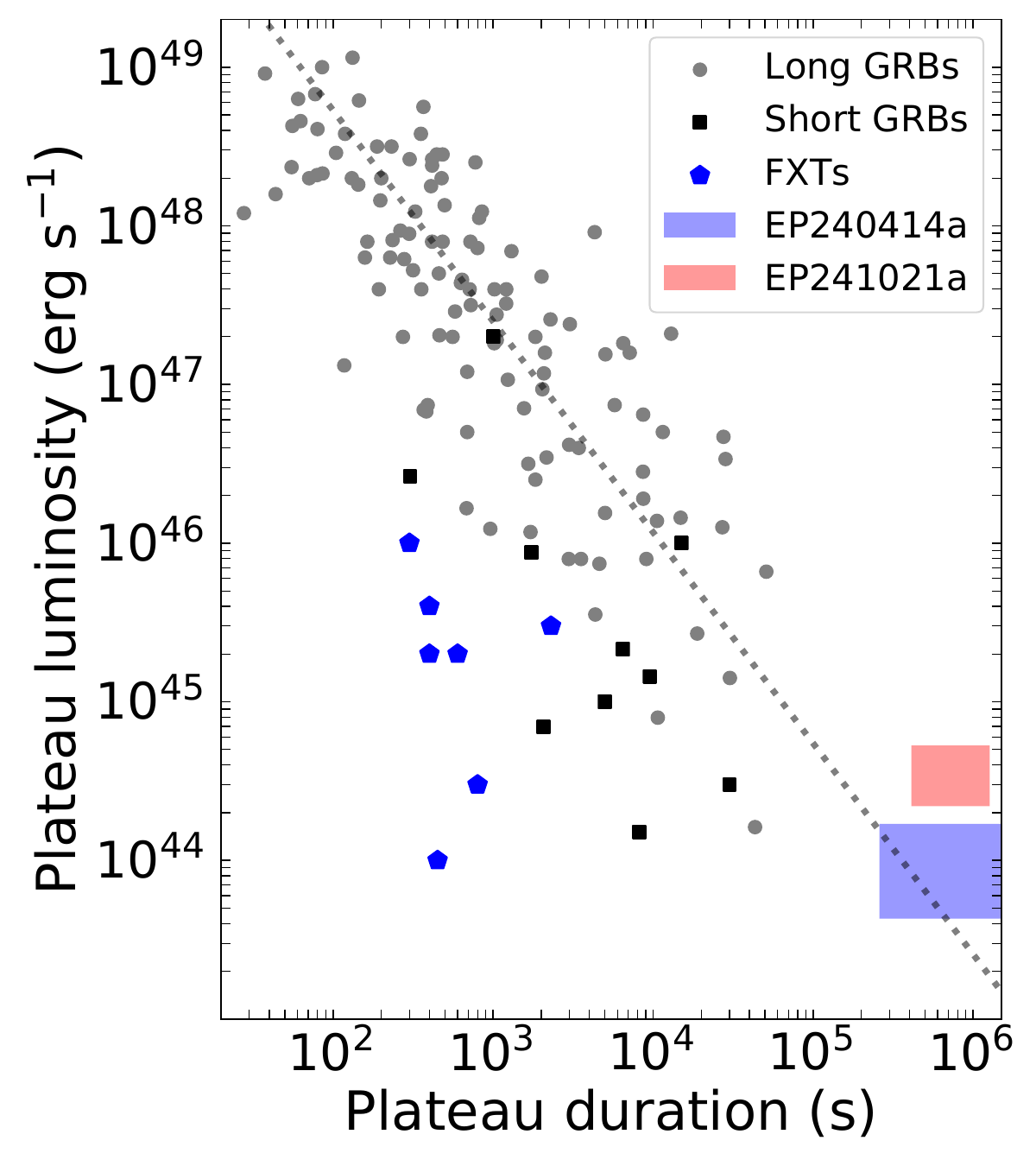}
    \caption{Observed X-ray plateau rest frame duration and luminosity for GRBs \citep{Tang2019,Xu2021} and FXTs  \citep{Quirola2024} versus EP240414a (blue shaded region; \citealt{Sun2024,vanDalen2024}) and EP241021a (red shaded region; \citealt{Shu2025}). The dotted line is shown to guide the eye and extend the GRB plateau correlation.  
    }
    \label{fig:plat}
\end{figure}

While the sparse \textit{Swift}/XRT lightcurve of EP241021a (\S \ref{sec:temporal} and Table \ref{tab: observationsXray}) displays a clearly shallow decay with $\alpha_X$\,$-0.5^{+0.4}_{-0.2}$, the refined EP/FXT lightcurve \citep{Gianfagna2025,Shu2025} further supports the long-lived shallow phase of the X-ray lightcurve. \citet{Shu2025} also finds a shallow decay of $\alpha_X$\,$=$\,$-0.31^{+0.17}_{-0.13}$ out to $6.1^{+8.6}_{-1.4}$ d (rest frame). While the exact mechanism producing the plateau is unclear, we can place its behavior in the context of the typical GRB X-ray plateau correlations. We find that the observed X-ray plateaus from EP240414a and EP241021a appear to be consistent with an extension of the typical rest-frame GRB X-ray plateau correlations between luminosity and duration (Figure~\ref{fig:plat}; e.g., \citealt{Dainotti2008,Tang2019,Xu2021}), which supports the interpretation of EP240414a and EP241021a in the context of GRBs. We display the allowed range of X-ray plateau luminosity and duration as shaded regions for both events. These shaded regions (Figure~\ref{fig:plat}) appear consistent with extrapolating the observed behavior of GRBs to lower luminosities and longer durations at the same level of scatter as the GRB plateaus in the more well populated region of parameter space, though see \citet{Shu2025} for an alternative interpretation\footnote{We note that the $L$-$T$-$E$ correlation \citep[e.g.,][]{Tang2019,Xu2021} utilized by \citet{Shu2025} depends on the uncertain isotropic-equivalent gamma-ray energy (see \S \ref{sec:prompt}). A larger isotopic energy (e.g., $10^{51-52}$ erg), which cannot be excluded by observations (Figure \ref{fig:gamma} and \S \ref{sec:trigger}), leads  EP241021 to be consistent with the GRB $L$-$T$-$E$ correlation \citep[Figure 8 of][]{Tang2019}. The correlation shown in Figure \ref{fig:plat} is independent of this uncertainty.}. These are among the longest plateaus detected from a GRB-like outflow \citep[e.g.,][]{Dainotti2008,Tang2019,Xu2021}. This motivates further late-time observations (e.g., \textit{Chandra} and \textit{XMM-Newton}) of EP sources, and improved early-time X-ray sampling by \textit{Swift} and EP.

\subsection{Constraints on the Initial Lorentz Factor}

\subsubsection{Lorentz Factor Limits from the Prompt Emission}

We find that the prompt emission spectrum observed by EP is non-thermal (see \S \ref{sec:prompt}) based on the inferred photon index $\Gamma$\,$=$\,$-1.80^{+0.57}_{-0.54}$ of the EP trigger   \citep{EP241021aWXT,Shu2025}.
Producing a non-thermal, likely synchrotron, spectrum requires that the energy is dissipated at a radius that is at least as large as the photosphere of the outflow \citep[e.g.,][]{Goodman1986,Thompson1994}.
Thus, we require that the emission occurs at a low optical depth $\tau$\,$\lesssim$\,$1$.
Based on the EP trigger, a minimum energy release is $3\times10^{49}$ erg ($0.5$\,$-$\,$4$ keV) over the $t_\textrm{dur}$\,$\sim$\,$100$ s trigger \citep{EP241021aWXT}, which is potentially much higher in larger energy bands.
In addition, the true isotropic kinetic energy $E_\textrm{kin}$ of the outflow is also likely much larger when accounting for radiation efficiency, and likely exceeds $E_\textrm{kin}$\,$\gtrsim$\,$10^{50}$ erg.
If the source is non-relativistic then $E_\textrm{kin}$\,$=$\,$E_\textrm{kin,true}$\,$\sim$\,$Mv^2/2$ and the radius as a function of time for a given outflow velocity $v$ is given by $R$\,$\sim$\,$vt$.
The optical depth of the outflow to Thomson scattering is then $\tau$\,$\sim$\,$2\kappa E_\textrm{kin}/(v^4t_\textrm{dur}^2)$ where $\kappa$ is the opacity.
As the optical depth decreases rapidly with an increase in velocity, we conservatively assume a non-relativistic outflow ($v$\,$\sim$\,$0.75c$, such that $\Gamma_0$\,$<$\,$1.5$ and relativistic effects can be ignored), which produces an extremely high optical depth of order hundreds. 
This would produce a thermal spectrum in sharp contrast to the observed prompt X-ray spectrum.

The optical depth can be decreased by assuming a higher velocity outflow.
Expanding this calculation to the transrelativistic regime, the optical depth is given by
\begin{align}
    \tau = 100 \frac{(\Gamma-1) E_{50}}{\Gamma_0^4\,\beta^2\, t_\textrm{dur,2}^2},
\end{align}
where we have applied the convention that $E_{50}$\,$=$\,$E/(10^{50}\, \textrm{erg})$ is the isotropic-equivalent energy and $t_\textrm{dur,2}$\,$=$\,$t_\textrm{dur}/(100\,\textrm{s})$ is the prompt emission duration.
By requiring $\tau$\,$<$\,$1$, we find an initial bulk Lorentz factor of $\Gamma_0$\,$\gtrsim$\,$4$.
If instead the isotropic-equivalent energy were higher, we would likewise require a larger Lorentz factor.
While we cannot rule out efficient dissipation in the mildly optically thick regime with at $\tau$ of a few, these calculations provide support for at least a mildly relativistic outflow to produce the prompt X-ray trigger of EP241021a. This is consistent with inferences for the Lorentz factor based on an analysis of the radio dataset \citep{Yadav2025, Shu2025} which likewise find  $\Gamma_0$\,$\gtrsim$\,$3$\,$-$\,$4$.

\subsubsection{Lorentz Factor Limits from Jet Deceleration}

If we assume the multi-wavelength emission from EP241021a is related to the interaction of a relativistic jet with its surrounding environment we can constrain the initial bulk Lorentz factor of the outflow.
For an on-axis observer, the afterglow emission will be observed to be declining after the jet has decelerated \citep[see, e.g.,][for a discussion of the impact of the viewing angle]{OConnor2020,OConnor2024}.
Therefore, based on the declining optical lightcurve starting at 1.14 d \citep[$0.65$ d in the rest frame;][]{2024GCN.37849....1Z}, we can set a lower limit to the initial bulk Lorentz factor $\Gamma_0$ of material at the jet's core.
A relativistic jet propagating into an external environment $\rho(r)$\,$\propto$\,$r^{-k}$, where $k$\,$=$\,$0$ represents a uniform density environment and $k$\,$=$\,$2$ for a stellar wind environment, begins to decelerate after a dynamical timescale $t_\mathrm{dec}$ given by \citep{Sari1999,Molinari2007,Ghisellini2010,Ghirlanda2012,Nava2013,Nappo2014,Ghirlanda2018}
\begin{align}
\label{eqn:tdec}
    \frac{t_\mathrm{dec}}{1+z}=\left(\frac{17-4k}{16\pi(4-k)}\frac{E_\textrm{kin}}{c^{5-k}\,A\, \Gamma_0^{8-2k} }\right)^{1/(3-k)},
\end{align}
where $\Gamma_0$ is the initial bulk Lorentz factor at the jet's core, $E_\mathrm{kin}$ is the kinetic energy, $A$\,$=$\,$m_\textrm{p}n_0 R_0^k$, $n_0$ is the density, $m_\textrm{p}$ is the proton mass, and $c$ is the speed of light. \autoref{eqn:tdec} simplifies to
\begin{align}
\label{eqn:tdecother}
    \frac{t_\textrm{dec}}{1+z} \lesssim \left\{ \begin{array}{ll}  0.68\, \Gamma_{0,1}^{-8/3}\,E_{\textrm{kin},52}^{1/3}\,n^{-1/3} \; \mathrm{d} & k=0, \\[2mm]
0.76\, \Gamma_{0,1}^{-4}\,E_{\textrm{kin},52}\,A_{*,-1}^{-1} \; \mathrm{d} & k=2,
\end{array} \right.
\end{align}
where $E_{\textrm{kin},52}$\,$=$\,$E_\textrm{kin}/(10^{52}\,\textrm{erg})$, $\Gamma_{0,1}$\,$=$\,$\Gamma_0$/10, and $A_*$\,$=$\,$n_0R_0^2/(3\times10^{35}\,\textrm{cm}^{-1})$ with $R_0$\,$=$\,$10^{18}$ cm such that $A_{*,-1}$\,$=$\,$A_*/0.1$.
The constraint on the Lorentz factor is degenerate with the kinetic energy of the blastwave and the density of the surrounding environment. If instead we assume a lower kinetic energy of $E_\textrm{kin}$\,$=$\,$10^{51}\,\textrm{erg}$ with $A_*$\,$=$\,$1$ and $n$\,$=$\,$1$ cm$^{-3}$, using Equation \ref{eqn:tdecother}, we require $\Gamma_0$\,$\gtrsim$\,$3.5$ $(7.5)$ for a wind environment and uniform density environment, respectively, in order for the jet to decelerate prior to the first optical observations.  Based on this calculation we conclude that the outflow has to be at least mildly relativistic. This conclusion holds over a broad range of kinetic energies between $E_\textrm{kin}$\,$=$\,$10^{50-53}\,\textrm{erg}$ and densities $A_*$\,$<$\,$10$ and $n$\,$<$\,$10$ cm$^{-3}$. Under these conditions, the required Lorentz factor to satisfy this condition is always $\Gamma_0$\,$\gtrsim$\,$1.5$ $(4)$ for a wind environment and uniform density environment, respectively. Lower values for the surrounding density or larger energies require significantly larger Lorentz factors than these quoted values, which are the requirements for the minimum quoted energy ($E_\textrm{kin}$\,$=$\,$10^{50}\,\textrm{erg}$) and the maximum quoted density (e.g., $n$\,$<$\,$10$ cm$^{-3}$).
It is important to note that there are no robust constraints on the exact start time of the jet's deceleration, and as such an extremely early deceleration (requiring higher Lorentz factors), typical of cosmological GRBs \citep[e.g.,][]{Ghirlanda2018}, cannot be excluded.

\subsection{Possible origins of EP241021a}
\label{sec:interp}

\subsubsection{A Rare Supernova}
\label{sec:mosfit}

\begin{figure}
    \centering
    \includegraphics[width=\columnwidth]{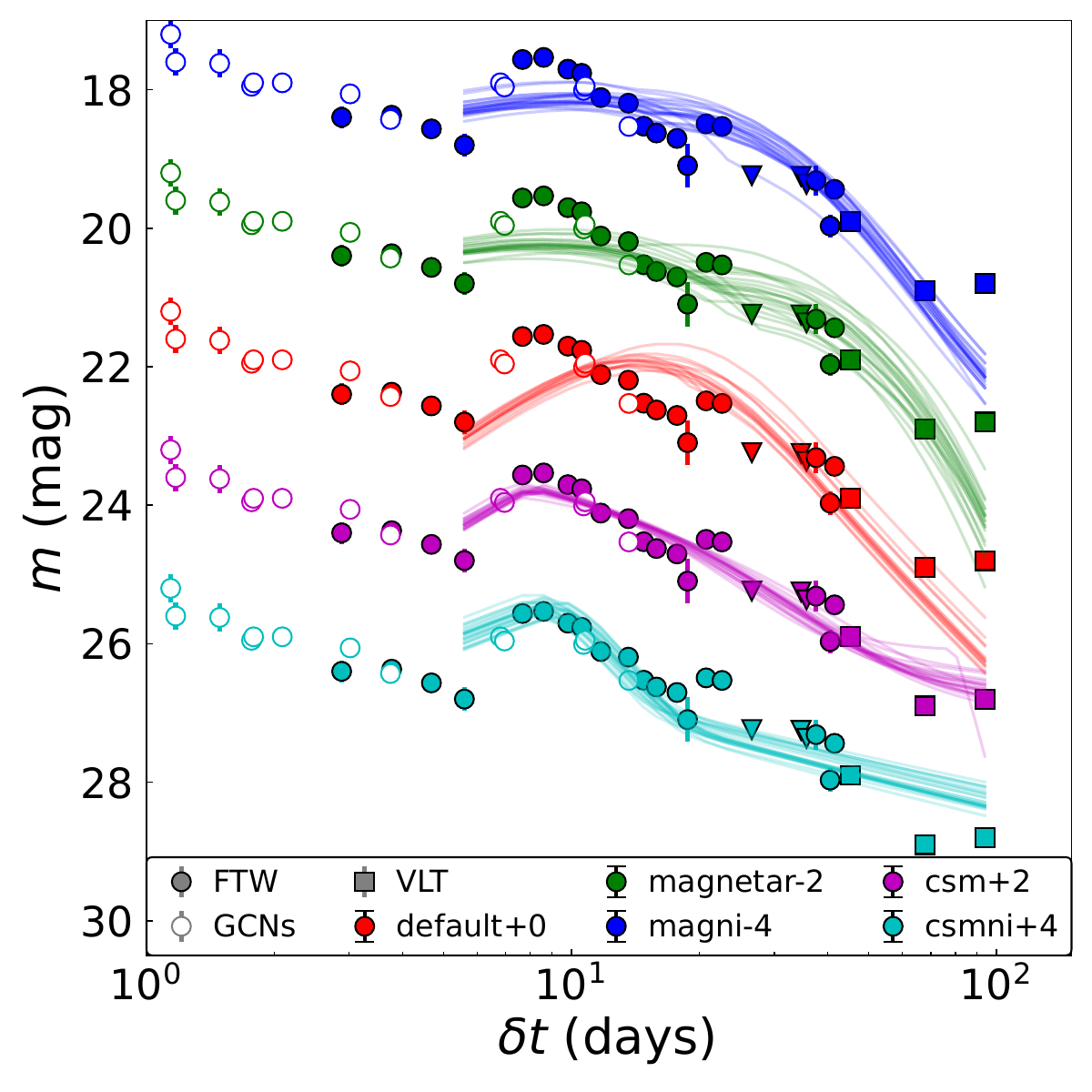}
    \caption{Comparison between supernova models and the observed optical lightcurve. The best fits to the second component are shown for a range of \texttt{MOSFiT} models (\default, \magnetar, \magni, \csm, and \csmni; see \S\ref{sec:mosfit}) in $r$-band. The model fits to the other filters ($z$ and $J$) are either quite poor, or require large amounts of dust and extreme absolute magnitudes.
    }
    \label{fig:mosfit}
\end{figure}

Motivated by the similarity of the absolute magnitude of the second bump to superluminous supernovae (SLSN), we used the Modular Open Source Fitter for Transients \citep[\mosfit;][]{Nicholl2017mosfit, Guillochon2018} to fit the rebrightening phase  ($>$\,$6$ d; observer frame) of EP241021a with supernova-like models. None of the five models (see Appendix  \ref{appendix:mosfit} for details) reproduced the light curve shape (Figure~\ref{fig:mosfit}), particularly the rapid evolution and early decay.

These models rely on thermal emission, which requires significant dust to match EP241021a’s spectrum ($F_\nu$\,$\propto$\,$\nu^{-1}$; Figure~\ref{fig:SEDs}). While the circumstellar material interaction models (\texttt{csmni} and \texttt{csm}; Appendix  \ref{appendix:mosfit}) offer a rough match to the post-peak behavior, they require extreme extinction (intrinsic $A_{V,z}$\,$\gtrsim$\,$1.5$ mag), implying unphysically high luminosities ($\lesssim$\,$-23.5$ to $-24.5$ mag), beyond observed SLSNe \citep[e.g.,][]{Gomez2024}.

Additional evidence against a SLSN origin includes EP241021a’s rapid evolution, red colors, and luminous non-thermal radio emission ($\sim$\,$10^{31}$ erg cm$^{-2}$ s$^{-1}$ Hz$^{-1}$; \citealt{Yadav2025,Gianfagna2025,Shu2025}; Srinivasaragavan et al., in preparation) -- several orders of magnitude above limits from known SLSNe \citep[e.g.,][]{Coppejans2018,Eftekhari2019}.

\subsubsection{An Off-axis Structured Jet}

\citet{BGG2020,BGG2022} showed that double-peaked afterglow lightcurves can be produced by a combination of line-of-sight material (producing the first peak) and the jet's core emission (producing the second peak).
The major factor dictating whether an observer views a single or double peaked lightcurve is the lowest latitude of the jet $\theta_*$, from which material is initially beamed towards the observer ($\theta_*\Gamma(\theta_*)$\,$=$\,$1$; see discussions in \citealt{BGG2020,BGG2022} and Figure~8 of \citealt{BGG2022}). If $\theta_\textrm{obs}$\,$>$\,$\theta_*$ then the the observer is continuously receiving emission from decreasing angles until finally observing the core, producing a single peaked lightcurve.
The alternative case, where $\theta_\textrm{obs}$\,$<$\,$\theta_*$ produces a double peaked lightcurve as initially the observer is viewing material far away from the core that will decelerate before the observer is able to receive emission from the core (due to relativistic beaming), producing a first peak at approximately $t_\textrm{dec}(\theta_\textrm{obs})$, see, e.g., \citet{OConnor2024}.
The secondary peak occurs once the core is de-beamed to the observer, and is the more standard peak for off-axis GRBs in the literature. 

The relative significance of the second peak is dictated by the slope of the structured jet with $E(\theta)$\,$\propto$\,$\theta^{-a}$.
Steeper jets and a uniform density environment serve to substantially pronounce the second peak (see Figure~11 of \citealt{BGG2022}). While this may be able to match the timescale and flux ratio of the double peaked OIR lightcurve of EP241021a, the observed rise to the second peak is too steep to naturally fit with this interpretation.
We therefore disfavor a standard forward shock from an off-axis structured jet as the sole mechanism for producing both the early and late OIR behavior observed in EP241021a. In order for the rebrightening to be due to an off-axis jet, the initial component would require a different emission mechanism \citep[e.g.,][]{Zheng2025,Gianfagna2025}. 


However, an alternative possibility is that the rebrightening is due to a reverse shock from off-axis material in a shallow structured jet \citep[e.g.,][]{OConnor2023,Gill2023,Zhang2024}.
This does not necessarily require an off-axis viewing angle for the observer.
Material at an off-axis angle would be moving with lower Lorentz factors and also be delayed with regard to photons reaching the observer (due to relativistic beaming).
This may provide a natural explanation for the observed delayed brightness increase in EP240414a and EP2141021a.
Recently \citet{Abdikamalov2025} performed a comprehensive investigation of the possible lightcurve behaviors of reverse and forward shock emission for steep off-axis jets.
They found a variety of possible double-peaked lightcurve profiles, due to de-beamed emission from the reverse shock.
Future work will extend these considerations to shallow structured jets, and may reveal the capability to reproduce the observed behavior of EP241021a due to emission from a reverse shock from an off-axis angle.

\subsubsection{A Refreshed Shock}
\label{sec:refresh}

Over the past two decades observations of GRBs have revealed a variety of peculiar, unexpected late-time behaviors.
These include long-lived X-ray \citep[e.g.,][]{Zhang2006,Troja2007} and optical \citep{Panaitescu2011,Knust2017} plateaus, late-time X-ray \citep{Burrows2005flare,Curran2008,Margutti2011,Bernardini2011} and optical \citep{Swenson2013,Kumar2022} flares, and rapid episodes of rebrightening \citep[e.g.,][]{Covino2008,Melandri2014,Dichiara2022,Moss2023}.
In many cases these phenomena have been attributed to long-lived central engine activity \citep{Burrows2005flare,Zhang2006,Curran2008,Margutti2011}, but other explanations are plausible \citep{Panaitescu2008,BeniaminiKumar2016,Beniamini2017,Lamberts2018,Ayache2020,Duque2022}.

Late-time rebrightening is often attributed to refreshed shocks \citep{ReesMeszaros1998,SariMeszaros2000,KumarPanaitescu2000,Johannesson2006}, such as observed in the afterglows of GRBs 030329 \citep{Granot2003refresh,Moss2023}, 071010A \citep{Covino2008}, GRB 120326A \citep{Melandri2014}, 
among others, though in some cases a reverse shock is also possible \citep[e.g.,][]{Dichiara2022}.
We note that extreme rebrightening episodes have also been observed in X-ray flashes (XRFs) such as XRF 050416a \citep{Soderberg2007XRF050416a} at $\sim$\,$25$ d (rest frame), which may also have been due to a refreshed shock.
In the standard refreshed shock model \citep{SariMeszaros2000,KumarPanaitescu2000},
promptly launched and rapid ($\Gamma_0$\,$\gtrsim$\,$100$) material decelerates as it sweeps up circumburst material in an external shock at the front of the outflow, eventually becoming slow enough for an initially less relativistic shell ($\Gamma_0$\,$\lesssim$\,$10$\,$-$\,$20$) to catch up and merge into a single blastwave, producing the increase in brightness, which is related to the energy added to the shock.

However, the rebrightening observed in EP241021a is at significantly later times ($\sim$\,$6.5\times10^5$ s; observer frame), and is difficult to obviously connect to the behavior observed in on-axis cosmological GRBs, especially in terms of the relative significance of the rebrightening (i.e., by more than a magnitude).
The time of the rebrightening can be related to the time of the shell collision when $\Gamma_f(t_\textrm{coll})$\,$=\Gamma_{0,s}/2$, which in a uniform density environment occurs at \citep{Moss2023}:
\begin{equation}
\label{eqn:tcoll}
   t_\textrm{coll} = 6.8\times10^5\, \textrm{s}\, \Bigg(\frac{1+z}{1.75}\Bigg)\, \Bigg(\frac{E_\textrm{kin}}{10^{53}\,\textrm{erg}}\Bigg)^{1/3} \Bigg(\frac{n}{1\,\textrm{cm}^{3}}\Bigg)^{-1/3} \Bigg(\frac{\Gamma_s}{6}\Bigg)^{-8/3}
\end{equation}
This timescale occurs for $\Gamma_s$\,$\approx$\,$6$ if the (isotropic-equivalent) kinetic energy is $\sim$\,$10^{53}$ erg, which is likely on the extreme end of the possible values.
If instead we consider a lower energy of $\sim$\,$10^{51}$ erg with fixed density $n$\,$=$\,$1$ cm$^{-3}$, the Lorentz factor of the slower moving material is then $\Gamma_s$\,$\approx$\,$3$ due to the weak dependence on energy in \autoref{eqn:tcoll}. 
Instead, in a wind environment, where $t_\textrm{coll}$\,$\propto$\,$(E_\textrm{kin}/A_*)\Gamma_s^{-4}$ \citep{Moss2023}, we find $\Gamma_s$\,$\approx$\,$7$\,$(2)$ for $E_\textrm{kin}$\,$\sim$\,$10^{53}$\,$(10^{51})$ erg.
Regardless, in either case, the trailing ejecta would have to at least be mildly relativistic, requiring that the initial outflow is also relativistic.

The rebrightening timescale of EP241021a is consistent with $\Delta t_\textrm{coll}/t_\textrm{coll}$\,$\lesssim$\,$0.3$, similar to the observations of GRB 030329 \citep{Granot2003refresh,Moss2023}. 
To good approximation, the rise time is the same as the reverse shock (RS) crossing time \citep{Sari1995}, which is determined by the ratio of Lorentz factors \citep[e.g,][]{Moss2023} such that $\Delta t_\textrm{coll}/t_\textrm{coll}$\,$\approx$\,$\Delta \Gamma_s/\Gamma_s$ to order unity corrections, where $\Delta \Gamma_s$ is the difference in Lorentz factors across the slower moving ejecta shell. 
We do not consider either a density enhancement \citep[e.g.,][]{Ramirez-Ruiz2001,Dai2002,Nakar2003,NakarGranot2007, vanEerten2009}, or complex density profile (potentially due to the progenitor star's mass loss history), or long-lived central engine activity to explain the rebrightening, as none of these explanations are capable of easily producing the variability timescale ($\Delta t/t$\,$\lesssim$\,$0.3$) and extremely steep rise required by the rebrightening of the OIR afterglow (steeper than $t^{3.5}$ at 90\% CL; see \S\ref{sec:temporal}).

The OIR temporal and spectral indices (\S\ref{sec:temporal} and \S\ref{sec:spectral}) are capable of matching standard afterglow closure relations for either a uniform density or wind environment \citep{Granot2002}. In a wind environment, as expected for a massive star progenitor, the X-ray, optical, and near-infrared data would be  above the cooling frequency ($\nu_\textrm{c}$\,$<$\,$\nu$) such that the temporal index and spectral index are $\alpha$\,$=$\,$(2-3p)/4$=$-1.15$ and $\beta$\,=$-p/2$\,$=$\,$-1.10$ for $p$\,$=$\,$2.2$ \citep{Granot2002}.
These indices are consistent with the values inferred in \S\ref{sec:temporal} and \S\ref{sec:spectral}.
We note that for emission above the cooling frequency the closure relation is unchanged between a uniform density or wind environment.
An additional possibility for a uniform density environment is a steep value of $p$\,$\approx$\,$3$ such that $\beta$\,=$(1-p)/2$\,$=$\,$-1$ and $\alpha$\,$=$\,$3(1-p)/4$=$-1.5$.
While the temporal slope has a better match for $p$\,$\approx$\,$2.5$, it is less consistent with the observed spectral index.
In either scenario, the radio data is below the injection frequency $\nu_\textrm{R}$\,$<$\,$\nu_\textrm{m}$\,$<$\,$\nu_\textrm{c}$ (\citealt{Yadav2025,Gianfagna2025,Shu2025}; Srinivasaragavan et al., in preparation).

The refreshed shock interpretation naturally explains the steep rise observed at OIR wavelengths (Figure~\ref{fig:optlcmag}), as well as the flat appearance of the poorly sampled \textit{Swift}/XRT X-ray lightcurve (Figure~\ref{fig:gamma}; right panel). We note that a refined EP/FXT lightcurve presented by \citet{Gianfagna2025,Shu2025} shows an X-ray bump that is simultaneous to the optical rebrightening. This is consistent with a refreshed shock interpretation, which should also impact the X-ray emission. In this scenario where the X-ray and optical behavior is linked to
the same emission mechanism or outflow, then the observed X-ray
lightcurve is not a true plateau (Figure \ref{fig:gamma}). However, \citet{Shu2025} note that the X-ray lightcurve has a plateau-like slope ($\alpha_X$\,$=$\,$-0.31^{+0.17}_{-0.13}$) out to $6.1^{+8.6}_{-1.4}$ d (rest frame), which we utilize in Figure \ref{fig:plat}. 
We also note that the radio lightcurves \citep{Yadav2025,Gianfagna2025,Shu2025} are not sampled at earlier times and all radio detections occur after the rebrightening phase, so a similar radio rebrightening is unconstrained.  
A refreshed relativistic shockwave also produces a non-thermal spectrum that provides a better description of the data than circumstellar medium interaction with non-relativistic supernova ejecta, which is commonly thought to produce a blackbody-like spectrum that is not observed (see Figure~\ref{fig:SEDs} and \S\ref{sec:mosfit}). 

\begin{figure}
\centering
\includegraphics[width=\columnwidth]{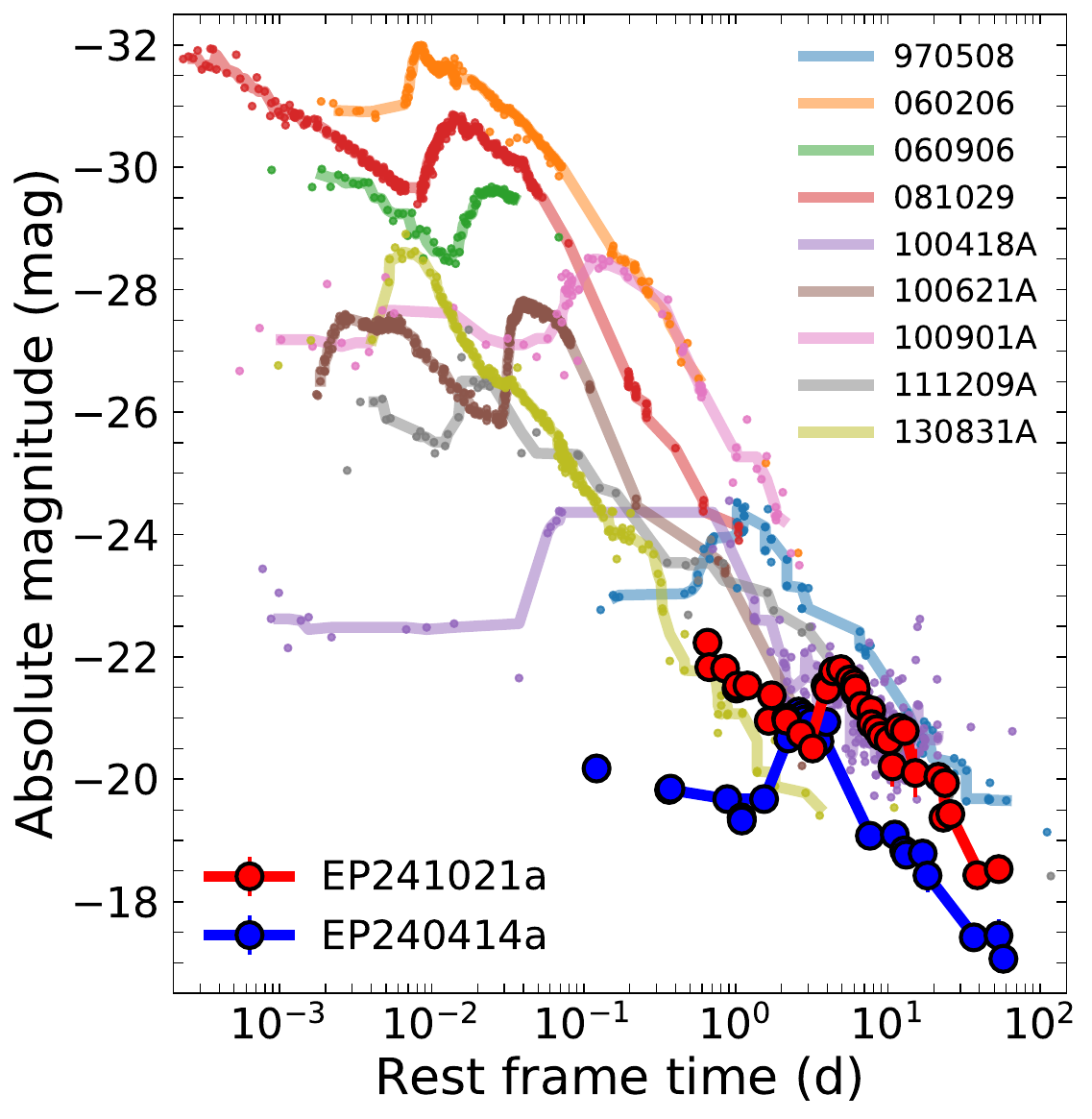}
\caption{Optical lightcurves ($r$-band) of gamma-ray bursts exhibiting a rapid and significant rebrightening (see \ref{sec:refresh}). These are compared to EP240414a \citep{Srivastav2024,vanDalen2024,Sun2024} and EP241021a (this work). The sample shown includes: GRBs 970508 \citep{Pian1998},  060206 \citep{Wozniak2006}, 060906 \citep{Cenko2009dark}, 081029 \citep{Nardini2011}, 100418A \citep{Marshall2011,deUgartePostigo2018}, 100621A \citep{Greiner2013}, 100901A \citep{Gorbovskoy2012}, 111209A \citep{Kann2018}, and 130831A \citep{Cano2014}. The solid lines show the data smoothed using a median filter. Reproduced from \citet{deUgartePostigo2018}. 
}
\label{fig:rebrighten}
\end{figure}

As our data do not show a clear jet-break out to late-times, we can set a constraint on the opening angle of the jet. We conservatively assume that the jet-break occurs after $t_\textrm{j}$\,$\gtrsim$\,$30$ d in the observer frame. 
Assuming an on-axis jet, the jet's half-opening angle is larger than \citep{SariPiranHalpern1999,Rhoads1999,Frail2001,ChevalierLi2000,Bloom2003}:
\begin{align}
    \theta_\textrm{c} = \left\{ \begin{array}{ll} 0.2\,\textrm{rad}\,\Bigg(\dfrac{t_\textrm{j}}{30\,\textrm{d}}\Bigg)^{3/8} \Bigg(\dfrac{E_\textrm{kin}}{10^{52}\,\textrm{erg}}\Bigg)^{-3/8} \Bigg(\dfrac{n}{10^{-2}\,\textrm{cm}^{-3}}\Bigg)^{3/8}, & k=0, \\[4mm]
0.13\,\textrm{rad}\,\Bigg(\dfrac{t_\textrm{j}}{30\,\textrm{d}}\Bigg)^{1/4} \Bigg(\dfrac{E_\textrm{kin}}{10^{52}\,\textrm{erg}}\Bigg)^{-1/4} \Bigg(\dfrac{A_*}{10^{-2}}\Bigg)^{1/4}, & k=2.
\end{array} \right.
\end{align}
where we have already applied the redshift  corrections using $z$\,$=$\,$0.748$. These half-opening angles are rather large for a GRB jet (which is not unexpected given the lack of a jet break to late times in EP241021a), especially for $k$\,$=$\,$0$ where $\theta_\textrm{c}$\,$\gtrsim$\,$10$ deg, but they are not completely unprecedented \citep[e.g.,][]{Frail2001,Wang2018,RoucoEscorial2022,OConnor2023}. However, different assumptions for the kinetic energy and density can decrease this opening angle slightly. We note that this is consistent with the inference of a wide-angled outflow by \citet{Yadav2025,Gianfagna2025} through an independent modeling of the broadband (X-ray to radio) dataset. As the trailing ejecta is likely more collimated than the initial outflow, a larger opening angle more easily favors the production of a rebrightening episode with $\Delta t_\textrm{coll}/t_\textrm{coll}$\,$\lesssim$\,$0.3$; see \citet{Moss2023} for a discussion.

Similarly significant and rapid optical rebrightenings with a change in magnitude of $\Delta m$\,$\gtrsim$\,$1$ mag have been found to occur in a handful of GRBs \citep[e.g., Figure 4 of][]{deUgartePostigo2018}, generally within a day after trigger (in the rest frame). The rebrightening of EP241021a is at the far end of this distribution of rebrightening times (see Table 4 of \citealt{deUgartePostigo2018}), but similar to GRB 970508 \citep{Pian1998}. In Figure \ref{fig:rebrighten}, we show a population of GRBs with extreme optical rebrightenings from the population of optical lightcurves compiled by \citet{Kann2006,Kann2010,Kann2011,Kann2018} and reproduced from \citet{deUgartePostigo2018}. The optical lightcurves shown in Figure \ref{fig:rebrighten} were compiled through the compilation\footnote{\url{https://grblc-catalog.streamlit.app/}} of \citet{Dainotti2024} and a public repository\footnote{\url{https://github.com/steveschulze/kann_optical_afterglows}} of Kann lightcurves \citep[see][]{Kann2006,Kann2010,Kann2011,Nicuesa2012}. We find that these events support our interpretation of EP241021a in the context of GRBs. 

Rapid afterglow rebrightening episodes such as those shown in Figure \ref{fig:rebrighten} are generally interpreted as continued energy injection from the central engine \citep[e.g.,][]{Marshall2011,Moin2013,Laskar2015}, though other possible interpretations exist and there is no strong consensus as to the exact mechanism. The continuous refreshing of the forward shock due to ejecta with a range of Lorentz factors\footnote{We point out that a refreshing of the forward shock by ejecta with a powerlaw distribution of Lorentz factors $M_\textrm{ej}(\Gamma)$\,$\propto$\,$\Gamma^{-s}$ \citep[e.g,][]{ReesMeszaros1998}, where $M_\textrm{ej}(\Gamma)$ is the ejecta mass for a given Lorentz factor, produces a different behavior compared to a discrete injection of energy (for a detailed discussion see \citealt{SariMeszaros2000}) as favored in this work.} \citep{Zhang2006,ReesMeszaros1998} is an alternative possibility \citep[for a discussion see, e.g.,][]{Marshall2011,Moin2013}. Other possible causes presented in the literature include a reverse shock \citep[e.g.,][]{Kobayashi2003,Covino2010}, a double jet \citep[e.g.,][]{Kann2018}, or an off-axis viewing angle \citep[e.g.,][]{Marshall2011,Greiner2013}.

As shown in Figure \ref{fig:rebrighten}, the rebrightening of EP241021a is at the extreme end of GRB optical luminosities (at a similar rest frame time), and the rebrightening phase is comparable to the luminosity of the brightest GRB afterglows, including the brightest-of-all-time GRB 221009A \citep[e.g.,][]{OConnor2023,Laskar2023}, at a similar (rest frame) time. Such extreme rebrightening behavior (Figure \ref{fig:rebrighten}) would be hard to miss in past GRBs, implying that EP240414a and EP241021a would potentially have unique central engine properties, possibly resulting from either a failed or marginally successful jet breakout, which can also explain the lack of gamma-rays. As EP240414a and EP241021a are clearly different from the majority of GRBs (though potentially showing similar behavior to a small minority; Figure \ref{fig:rebrighten}, \citealt{deUgartePostigo2018}), it is possible that other properties of their outflows are atypical, such as their Lorentz factors being mildly relativistic and the presence of trailing ejecta with sufficient energy to lead to a significant bump.
Further exploration of the detectability of similar refreshed shocks in typical cosmological GRBs is left for future work, and will aid in determining the uniqueness of this refreshed shock model for EP241021a.

\section{Conclusions}

We have presented the results of our multi-wavelength follow-up campaign of EP241021a over the first 100 days of its evolution, including long-term monitoring with the FTW, VLT, and HET. Here, we summarize our main conclusions:
   \begin{itemize}
      \item The non-thermal prompt soft X-ray emission detected by EP requires at least a mildly relativistic outflow with bulk Lorentz factor $\Gamma$\,$\gtrsim$\,$4$.
      This interpretation is supported by the multiple independent analyses of the comprehensive radio dataset, which likewise require a mildly relativistic outflow  \citep[see][]{Yadav2025,Shu2025}.
      \item The rapid optical rebrightening (Figure~\ref{fig:optlcmag}) is challenging to explain with supernova-like models (Figure~\ref{fig:mosfit}), but may be related to either refreshed shocks or a reverse shock from off-axis material. In either scenario we favor an at least mildly relativistic outflow producing synchrotron radiation.
   \end{itemize}

Future observations of similar events are required to build a sample, which will aid in determining the progenitor system, emission mechanisms, and diversity of their outflows. We note that a refined X-ray dataset (i.e., higher cadence) and multi-wavelength observations at $<$\,$1$ d would provide stronger constraints on the available models and the onset of the first component and its Lorentz factor.

\begin{acknowledgements}
The authors thank the referee for their careful review and comments that improved the manuscript. 
B. O. acknowledges useful discussions with Dheeraj Pasham, Chris Irwin, Zorawar Wadiasingh, and Jonathan Granot. 
M. B. is supported by a Student Grant from the Wübben Stiftung Wissenschaft.
B. O. is supported by the McWilliams Postdoctoral Fellowship at Carnegie Mellon University.
P. B. is supported by a grant (no.
2020747) from the United States-Israel Binational Science Foundation (BSF), Jerusalem, Israel, by a grant (no.
1649/23) from the Israel Science Foundation and by a grant (no.
80NSSC 24K0770) from the NASA astrophysics theory program.
M. M.'s research was supported by an appointment to the NASA Postdoctoral Program at the NASA Goddard Space Flight Center, administered by Oak Ridge Associated Universities under contract with NASA\@.
E. T., Y.-H. Y., and M. Y. are supported by the European Research Council through the Consolidator grant BHianca (grant agreement ID 101002761).
This paper contains data obtained at the Wendelstein Observatory of the Ludwig-Maximilians University Munich.
Funded by the Deutsche Forschungsgemeinschaft (DFG, German Research Foundation) under Germany's Excellence Strategy – EXC-2094 – 390783311.
This research has made use of the XRT Data Analysis Software (XRTDAS) developed under the responsibility of the ASI Science Data Center (ASDC), Italy.
This research has made use of data and/or software provided by the High Energy Astrophysics Science Archive Research Center (HEASARC), which is a service of the Astrophysics Science Division at NASA/GSFC\@.
This paper contains data from observations obtained with the Hobby-Eberly Telescope (HET), which is a joint project of the University of Texas at Austin, the Pennsylvania State University, Ludwig-Maximillians-Universität München, and Georg-August Universität Göttingen.
The HET is named in honor of its principal benefactors, William P. Hobby and Robert E. Eberly.
We thank Cassie Crowe, Nathan McReynolds, Stephen Odewahn, Justen Pautzke, Amy Ray, Sergey Rostopchin and Amy Westfall from the HET staff for obtaining the observations.
We acknowledge the Texas Advanced Computing Center (TACC) at The University of Texas at Austin for providing high performance computing, visualization, and storage resources that have contributed to the results reported within this paper.
The Low Resolution Spectrograph 2 (LRS2) was developed and funded by the University of Texas at Austin McDonald Observatory and Department of Astronomy, and by Pennsylvania State University.
We thank the Leibniz-Institut fur Astrophysik Potsdam (AIP) and the Institut fur Astrophysik Goettingen (IAG) for their contributions to the construction of the integral field units.
This research has made use of the Astrophysics Data System, funded by NASA under Cooperative Agreement 80NSSC21M00561.
This research has made use of adstex (\url{https://github.com/yymao/adstex}).
This work made use of the catalog presented by \citet{Dainotti2024}. This work made use of the Astro-COLIBRI platform \citep{2021ApJS..256....5R}.
\end{acknowledgements}

%
%

\bibliographystyle{aa} 
\bibliography{epbib}

\onecolumn
\begin{appendix}
    \section{Log of Observations}

    Here we report the list of observations of EP241021a analyzed as part of this work.
    The observations are tabulated in Table~\ref{tab: observationsPhot}, \ref{tab: observationsSpec}, and \ref{tab: observationsXray}. 

\begin{longtable}{lS[table-format=2.2]ccS[table-format=5]cS[table-format=2.2\pm0.2]S[table-format=0.2]}
    \caption{Log of optical and near-infrared imaging observations of EP241021a used in this work. The photometry is not corrected for Galactic extinction $E(B-V)=0.045$ mag \citep{Schlafly2011}. The Galactic extinction in a given bandpass is tabulated below as $A_\lambda$ \citep{Schlafly2011}.}
    \label{tab: observationsPhot} \\

    \toprule
    \textbf{Start Time (UT)} & \textbf{$\delta T$ (d)} & \textbf{Telescope} & \textbf{Instrument} & \textbf{Exposure (s)} & \textbf{Filter} & \textbf{AB magnitude} & \textbf{$A_\lambda$ (mag)} \\
    \midrule
    \endfirsthead

    \caption[]{(continued)} \\
    \toprule
    \textbf{Start Time (UT)} & \textbf{$\delta T$ (d)} & \textbf{Telescope} & \textbf{Instrument} & \textbf{Exposure (s)} & \textbf{Filter} & \textbf{AB magnitude} & \textbf{$A_\lambda$ (mag)} \\
    \midrule
    \endhead

    \midrule
    \multicolumn{8}{r}{\textit{Continued on next page}} \\
    \endfoot

    \bottomrule
    \endlastfoot

    \multicolumn{8}{c}{\textbf{Wendelstein Observatory}} \\
    2024-10-24T02:12:17      & 2.88                    & FTW                & 3KK                 & 1980                  & \textit{r}      & 22.48\pm0.16           & 0.10                       \\
    2024-10-24T02:12:17      & 2.88                    & FTW                & 3KK                 & 1980                  & \textit{z}      & 21.55\pm0.21           & 0.06                       \\
    2024-10-24T02:12:31      & 2.88                    & FTW                & 3KK                 & 1867                  & \textit{J}      & 21.04\pm0.21           & 0.04                       \\
    2024-10-24T23:42:00      & 3.77                    & FTW                & 3KK                 & 7200                  & \textit{r}      & 22.47\pm0.10           & 0.10                       \\
    2024-10-24T23:42:00      & 3.77                    & FTW                & 3KK                 & 7200                  & \textit{z}      & 22.20\pm0.12           & 0.06                       \\
    2024-10-24T23:42:14      & 3.77                    & FTW                & 3KK                 & 6790                  & \textit{J}      & 21.73\pm0.12           & 0.04                       \\
    2024-10-25T21:27:05      & 4.68                    & FTW                & 3KK                 & 9540                  & \textit{r}      & 22.69\pm0.10           & 0.10                       \\
    2024-10-25T21:27:05      & 4.68                    & FTW                & 3KK                 & 9540                  & \textit{z}      & 22.50\pm0.21           & 0.06                       \\
    2024-10-25T21:27:20      & 4.68                    & FTW                & 3KK                 & 8147                  & \textit{J}      & 22.08\pm0.23           & 0.04                       \\
    2024-10-26T19:25:35      & 5.60                    & FTW                & 3KK                 & 5580                  & \textit{r}      & 22.93\pm0.18           & 0.10                       \\
    2024-10-26T19:25:35      & 5.60                    & FTW                & 3KK                 & 5400                  & \textit{z}      & {>22.2}               & 0.06                       \\
    2024-10-26T19:25:49      & 5.60                    & FTW                & 3KK                 & 5262                  & \textit{J}      & 21.44\pm0.26           & 0.04                       \\
    2024-10-28T21:04:27      & 7.66                    & FTW                & 3KK                 & 5580                  & \textit{r}      & 21.67\pm0.10           & 0.10                       \\
    2024-10-28T21:04:27      & 7.66                    & FTW                & 3KK                 & 5580                  & \textit{z}      & 21.20\pm0.10           & 0.06                       \\
    2024-10-28T21:08:46      & 7.67                    & FTW                & 3KK                 & 5092                  & \textit{J}      & 20.79\pm0.10           & 0.04                       \\
    2024-10-29T19:23:45      & 8.59                    & FTW                & 3KK                 & 4500                  & \textit{r}      & 21.63\pm0.10           & 0.10                       \\
    2024-10-29T19:23:45      & 8.59                    & FTW                & 3KK                 & 4500                  & \textit{z}      & 21.09\pm0.10           & 0.06                       \\
    2024-10-29T19:23:59      & 8.59                    & FTW                & 3KK                 & 3904                  & \textit{J}      & 20.77\pm0.10           & 0.04                       \\
    2024-10-31T00:20:10      & 9.80                    & FTW                & 3KK                 & 5040                  & \textit{r}      & 21.80\pm0.10           & 0.10                       \\
    2024-10-31T00:20:10      & 9.80                    & FTW                & 3KK                 & 5040                  & \textit{z}      & 21.32\pm0.10           & 0.06                       \\
    2024-10-31T00:20:24      & 9.80                    & FTW                & 3KK                 & 4753                  & \textit{J}      & 20.92\pm0.10           & 0.04                       \\
    2024-10-31T18:40:04      & 10.56                   & FTW                & 3KK                 & 7200                  & \textit{r}      & 21.87\pm0.10           & 0.10                       \\
    2024-10-31T18:40:04      & 10.56                   & FTW                & 3KK                 & 7200                  & \textit{z}      & 21.47\pm0.10           & 0.06                       \\
    2024-10-31T18:40:18      & 10.56                   & FTW                & 3KK                 & 6450                  & \textit{J}      & 21.12\pm0.10           & 0.04                       \\
    2024-11-01T22:17:50      & 11.72                   & FTW                & 3KK                 & 7200                  & \textit{r}      & 22.23\pm0.10           & 0.10                       \\
    2024-11-01T22:17:50      & 11.72                   & FTW                & 3KK                 & 7020                  & \textit{z}      & 21.66\pm0.10           & 0.06                       \\
    2024-11-01T22:18:04      & 11.72                   & FTW                & 3KK                 & 6790                  & \textit{J}      & 21.24\pm0.10           & 0.04                       \\
    2024-11-03T19:34:27      & 13.60                   & FTW                & 3KK                 & 7200                  & \textit{r}      & 22.30\pm0.10           & 0.10                       \\
    2024-11-03T19:34:27      & 13.60                   & FTW                & 3KK                 & 7200                  & \textit{z}      & 21.84\pm0.12           & 0.06                       \\
    2024-11-03T19:34:41      & 13.60                   & FTW                & 3KK                 & 6790                  & \textit{J}      & 21.90\pm0.20           & 0.04                       \\
    2024-11-04T23:53:42      & 14.78                   & FTW                & 3KK                 & 6480                  & \textit{r}      & 22.61\pm0.10           & 0.10                       \\
    2024-11-04T23:53:42      & 14.78                   & FTW                & 3KK                 & 6480                  & \textit{z}      & 21.80\pm0.10           & 0.06                       \\
    2024-11-04T23:53:56      & 14.78                   & FTW                & 3KK                 & 6111                  & \textit{J}      & 21.78\pm0.14           & 0.04                       \\
    2024-11-06T01:11:18      & 15.84                   & FTW                & 3KK                 & 3780                  & \textit{r}      & 22.73\pm0.10           & 0.10                       \\
    2024-11-06T01:11:18      & 15.84                   & FTW                & 3KK                 & 3780                  & \textit{z}      & 22.00\pm0.14           & 0.06                       \\
    2024-11-06T01:11:32      & 15.84                   & FTW                & 3KK                 & 3395                  & \textit{J}      & 21.87\pm0.20           & 0.04                       \\
    2024-11-07T21:53:13      & 17.70                   & FTW                & 3KK                 & 7200                  & \textit{r}      & 22.80\pm0.10           & 0.10                       \\
    2024-11-07T21:53:13      & 17.70                   & FTW                & 3KK                 & 7200                  & \textit{z}      & 21.97\pm0.11           & 0.06                       \\
    2024-11-07T21:53:27      & 17.70                   & FTW                & 3KK                 & 6790                  & \textit{J}      & 21.93\pm0.16           & 0.04                       \\
    2024-11-08T22:55:09      & 18.74                   & FTW                & 3KK                 & 4860                  & \textit{r}      & 23.2 \pm0.3            & 0.10                       \\
    2024-11-08T22:55:09      & 18.74                   & FTW                & 3KK                 & 4680                  & \textit{z}      & {>22.1}               & 0.06                       \\
    2024-11-08T23:02:27      & 18.75                   & FTW                & 3KK                 & 1188                  & \textit{J}      & {>21.2}               & 0.04                       \\
    2024-11-10T22:29:33      & 20.72                   & FTW                & 3KK                 & 1800                  & \textit{r}      & 22.59\pm0.11           & 0.10                       \\
    2024-11-10T22:29:33      & 20.72                   & FTW                & 3KK                 & 1800                  & \textit{z}      & 22.26\pm0.22           & 0.06                       \\
    2024-11-10T22:29:47      & 20.72                   & FTW                & 3KK                 & 1528                  & \textit{J}      & {>22.3}               & 0.04                       \\
    2024-11-12T19:22:13      & 22.59                   & FTW                & 3KK                 & 5400                  & \textit{r}      & 22.64\pm0.12           & 0.10                       \\
    2024-11-12T19:22:13      & 22.59                   & FTW                & 3KK                 & 5400                  & \textit{z}      & 22.15\pm0.16           & 0.06                       \\
    2024-11-12T19:22:28      & 22.59                   & FTW                & 3KK                 & 5092                  & \textit{J}      & 22.28\pm0.22           & 0.04                       \\
    2024-11-16T18:12:12      & 26.54                   & FTW                & 3KK                 & 4140                  & \textit{r}      & 23.3 \pm0.4            & 0.10                       \\
    2024-11-16T18:12:12      & 26.54                   & FTW                & 3KK                 & 3960                  & \textit{z}      & 22.7 \pm0.4            & 0.06                       \\
    2024-11-16T18:12:26      & 26.54                   & FTW                & 3KK                 & 3904                  & \textit{J}      & {>22.5}               & 0.04                       \\
    2024-11-24T20:59:23      & 34.66                   & FTW                & 3KK                 & 9180                  & \textit{r}      & {>23.3}               & 0.10                       \\
    2024-11-24T20:59:23      & 34.66                   & FTW                & 3KK                 & 9360                  & \textit{z}      & {>22.1}               & 0.06                       \\
    2024-11-24T20:59:37      & 34.66                   & FTW                & 3KK                 & 8703                  & \textit{J}      & {>21.7}               & 0.04                       \\
    2024-11-25T21:53:13      & 35.70                   & FTW                & 3KK                 & 3600                  & \textit{r}      & {>23.5}               & 0.10                       \\
    2024-11-25T21:53:13      & 35.70                   & FTW                & 3KK                 & 3600                  & \textit{z}      & {>21.9}               & 0.06                       \\
    2024-11-25T21:53:27      & 35.70                   & FTW                & 3KK                 & 3395                  & \textit{J}      & {>21.3}               & 0.04                       \\
    2024-11-27T19:31:59      & 37.60                   & FTW                & 3KK                 & 4320                  & \textit{r}      & 23.39\pm0.23           & 0.10                       \\
    2024-11-27T19:31:59      & 37.60                   & FTW                & 3KK                 & 3420                  & \textit{z}      & {>22.5}               & 0.06                       \\
    2024-11-27T19:32:13      & 37.60                   & FTW                & 3KK                 & 3565                  & \textit{J}      & {>21.9}               & 0.04                       \\
    2024-11-30T18:51:08      & 40.57                   & FTW                & 3KK                 & 14220                 & \textit{r}      & 24.06\pm0.19           & 0.10                       \\
    2024-11-30T18:51:08      & 40.57                   & FTW                & 3KK                 & 14400                 & \textit{z}      & {>23.3}               & 0.06                       \\
    2024-11-30T18:51:22      & 40.57                   & FTW                & 3KK                 & 13579                 & \textit{J}      & 22.46\pm0.23           & 0.04                       \\
    2024-12-01T18:26:24      & 41.55                   & FTW                & 3KK                 & 10850                 & \textit{r}      & 23.49\pm0.12           & 0.10                       \\
    2024-12-01T18:26:24      & 41.55                   & FTW                & 3KK                 & 10850                 & \textit{i}      & 23.35\pm0.15           & 0.08                       \\
    2024-12-01T18:26:38      & 41.55                   & FTW                & 3KK                 & 9145                  & \textit{H}      & {>22.3}               & 0.02                       \\
    \midrule
    \multicolumn{8}{c}{\textbf{Very Large Telescope}} \\
    2024-12-05T01:48:40      & 45.2                    & VLT                & FORS2               & 1200                  & \textit{R}      & 23.9\pm0.1             & 0.12                       \\
    2024-12-05T02:13:04      & 45.2                    & VLT                & HAWK-I              & 900                   & \textit{J}      & 23.0\pm0.1             & 0.04                       \\
    2024-12-22T01:02:41      & 61.8                    & VLT                & HAWK-I              & 900                   & \textit{J}      & 23.3\pm0.1             & 0.04                       \\
    2024-12-28T02:12:42      & 67.9                    & VLT                & FORS2               & 1200                  & \textit{R}      & 24.90\pm0.15           & 0.12                       \\
    2025-01-23T01:04:23      & 93.8                    & VLT                & HAWK-I              & 1200                  & \textit{J}      & 24.02\pm0.15           & 0.04                       \\
    2025-01-23T01:32:49      & 93.8                    & VLT                & FORS2               & 1200                  & \textit{R}      & 24.8\pm0.1             & 0.12                       \\
    \midrule
    \multicolumn{8}{c}{\textbf{\textit{Swift}/UVOT}} \\
    2024-10-24T20:58:13      & 3.66                    & \textit{Swift}     & UVOT                & 2545                  & \textit{v}      & {>20.5}               & 0.14                       \\
    2024-10-24T21:08:18      & 3.67                    & \textit{Swift}     & UVOT                & 163                   & \textit{b}      & {>19.8}               & 0.17                       \\
    2024-10-25T15:30:30      & 4.43                    & \textit{Swift}     & UVOT                & 1347                  & \textit{uvm2}   & {>22.4}               & 0.39                       \\
    2024-10-29T09:32:49      & 8.18                    & \textit{Swift}     & UVOT                & 4289                  & \textit{u}      & 21.85\pm0.23           & 0.22                       \\
    2024-11-03T02:25:38      & 12.88                   & \textit{Swift}     & UVOT                & 3059                  & \textit{u}      & {>22.3}               & 0.22                       \\
    2024-11-05T20:24:14      & 15.63                   & \textit{Swift}     & UVOT                & 985                   & \textit{u}      & {>21.7}               & 0.22                       \\
    2024-11-07T02:15:17      & 16.88                   & \textit{Swift}     & UVOT                & 3107                  & \textit{u}      & {>22.3}               & 0.22                       \\
    2024-11-15T19:08:48      & 25.58                   & \textit{Swift}     & UVOT                & 1676                  & \textit{u}      & {>22.0}               & 0.22                       \\
\end{longtable}

\begin{table*}[ht]
    \centering
    \caption{Log of optical spectroscopic observations of EP241021a used in this work.
    }
    \label{tab: observationsSpec}
    \begin{tabular}{lS[table-format=2.2]ccS[table-format=4]}
        \toprule
        \textbf{Start Time (UT)} & \textbf{$\delta T$ (d)} & \textbf{Telescope} & \textbf{Instrument} & \textbf{Exposure (s)}  \\
        \midrule
        2024-10-25T04:25         & 5.0                     & HET                & LRS-B               & 3000                  \\
        2024-10-29T07:43         & 9.1                     & HET                & LRS-R               & 2050                  \\
        2024-11-26T05:43         & 37.0                    & HET                & LRS-R               & 3600                  \\
        \bottomrule
    \end{tabular}
\end{table*}

\begin{table*}[ht]
    \centering
    \caption{Log of X-ray observations of EP241021a used in this work. The unabsorbed flux is given in the $0.3-10$ keV band. Additional X-ray observations of EP241021a are reported by \citet{Shu2025,Gianfagna2025,Yadav2025}.
    }
    \label{tab: observationsXray}
    \begin{tabular}{lS[table-format=2.2]cccccc}
        \toprule
        \textbf{Start Time (UT)} & \textbf{$\delta T$ (d)} & \textbf{Telescope} & \textbf{Instrument} & \textbf{Exposure (s)} & \textbf{Flux (erg cm$^{-2}$ s$^{-1}$)}  \\[0.5mm]
        \midrule
        2024-10-24T20:54:12      & 3.66                    & \textit{Swift}     & XRT                 & 2770                  & $<2.5\times 10^{-13}$ \\[0.5mm]
        2024-10-25T15:26:02      & 4.43                    & \textit{Swift}     & XRT                 & 1380                  & $<3.9\times 10^{-13}$\\[0.5mm]
        2024-10-29T09:30:13      & 8.18                    & \textit{Swift}     & XRT                 & 4385                  & $(1.4^{+0.6}_{-0.5})\times 10^{-13}$ \\[0.5mm]
        2024-11-03T02:22:02      & 12.88                   & \textit{Swift}     & XRT                 & 3120                  & $<1.8\times 10^{-13}$ \\[0.5mm]
        2024-11-05T20:21:02      & 15.63                   & \textit{Swift}     & XRT                 & 1005                  & $<5.6\times 10^{-13}$ \\[0.5mm]
        2024-11-07T02:13:02      & 16.88                   & \textit{Swift}     & XRT                 & 3180                  & $(5.5^{+0.3}_{-0.3})\times 10^{-14}$ \\[0.5mm]
        2024-11-15T19:06:02      & 25.58                   & \textit{Swift}     & XRT                 & 1710                  & $(7.9^{+0.6}_{-0.4})\times 10^{-14}$ \\[0.5mm]
        \bottomrule
    \end{tabular}
\end{table*}

\section{Additional Results of OIR Lightcurve Fitting}
\label{Appendix Fit Results}

In \S \ref{sec:temporal}, we outline our temporal and spectral fit to the OIR lightcurve of EP241021a using Equation \ref{eqn:lc}. In Figure \ref{fig:optlcmag} we highlight the best fit to the lightcurve, where the initial $\alpha_1$ and final $\alpha_3$ temporal slopes are allowed to be independent. This provides an improved $\chi^2$ over the more restrictive fit where we require $\alpha_1$\,$=$\,$\alpha_3$. The fit to the OIR lightcurves and the fit residuals for $\alpha_1$\,$=$\,$\alpha_3$ are shown in Figure \ref{fig:optlcmag-same-slopes}, see also \S \ref{sec:temporal} for a more detailed discussion. Due to the worse residuals when $\alpha_1$\,$=$\,$\alpha_3$ we favor the fit shown in Figure \ref{fig:optlcmag}.

\begin{figure}
    \centering
    \includegraphics[width=\textwidth]{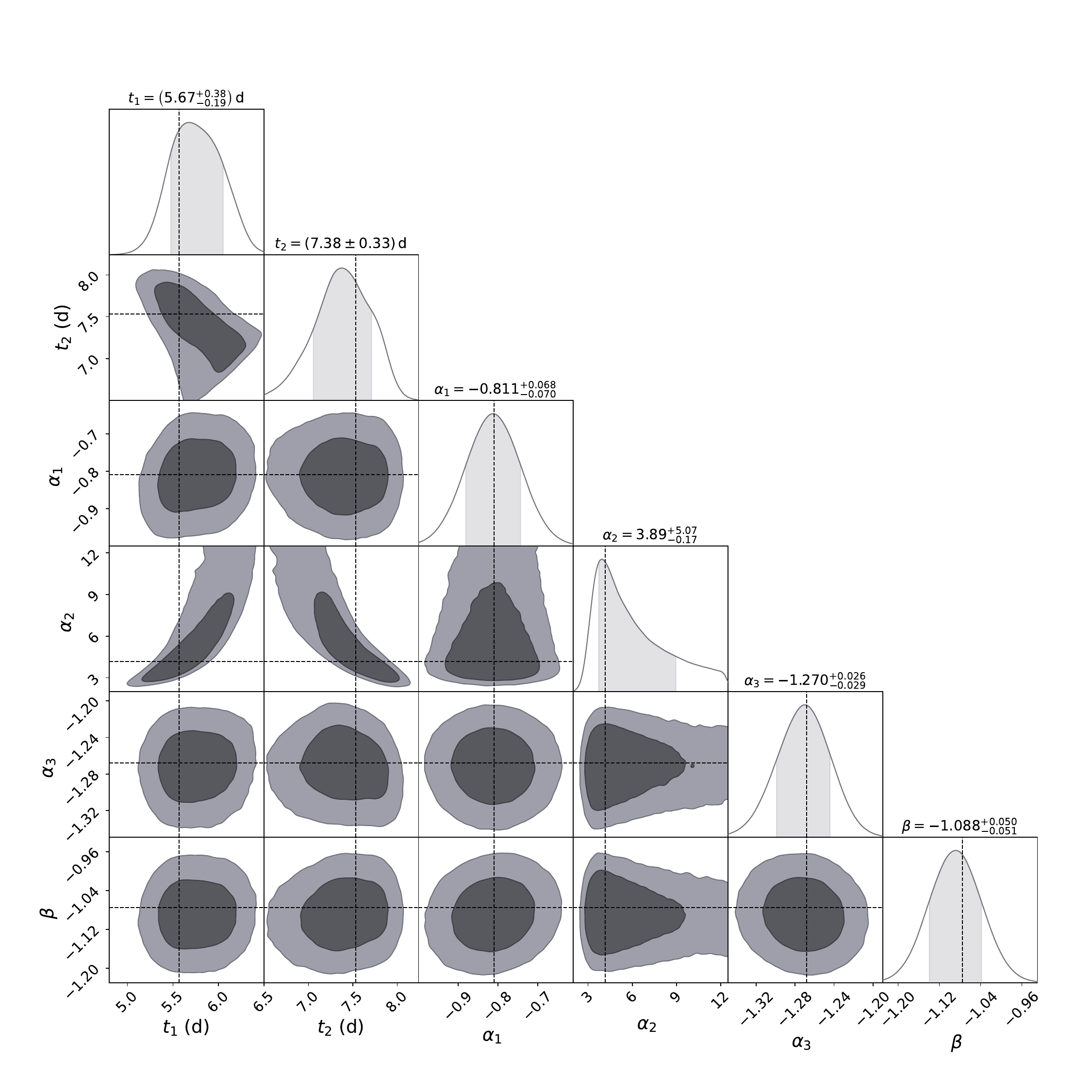}
    \caption{Corner plot for the broken powerlaw temporal and spectral fit $F_\nu$\,$\propto$\,$t^\alpha\nu^\beta$ using Equation \ref{eqn:lc} to the multi-band $rzJ$ lightcurves of EP241021a (see \S\ref{sec:temporal}).
    The dashed lines indicate the best fit values in the full chain. The fit has $\chi^2 / \mathrm{dof} = 164 / 77 = 2.13$.
    }
    \label{fig:power-laws-corner}
\end{figure}

\begin{figure*}
    \centering
\includegraphics[width=\textwidth]{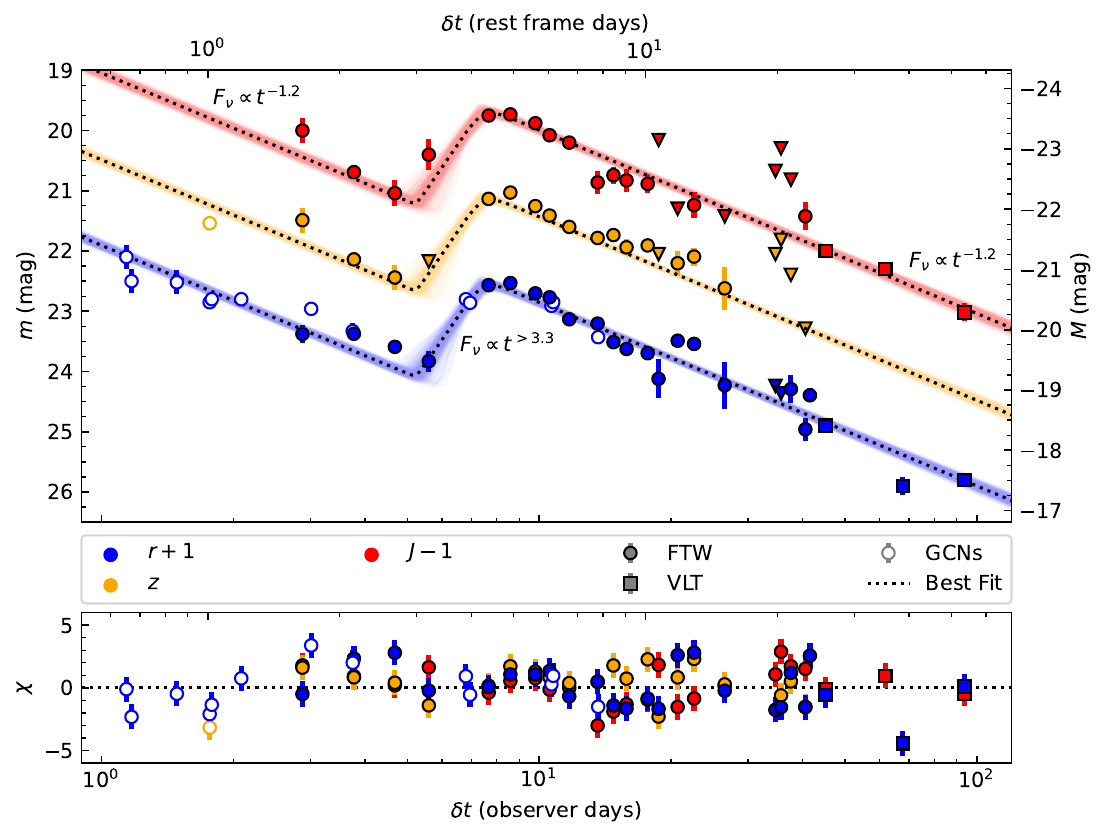}
\caption{Same as Figure \ref{fig:optlcmag} but for the temporal and spectral fit (see Equation \ref{eqn:lc}) where $\alpha_1$\,$=$\,$\alpha_3$. \textbf{Top:} The fit to the multi-filter lightcurve. 
\textbf{Bottom:} The fit residuals are shown in each filter. The residuals are more significant than in Figure \ref{fig:optlcmag}, especially at early times in $r$-band.
}
    \label{fig:optlcmag-same-slopes}
\end{figure*}

\begin{figure}
    \centering
    \includegraphics[width=\textwidth]{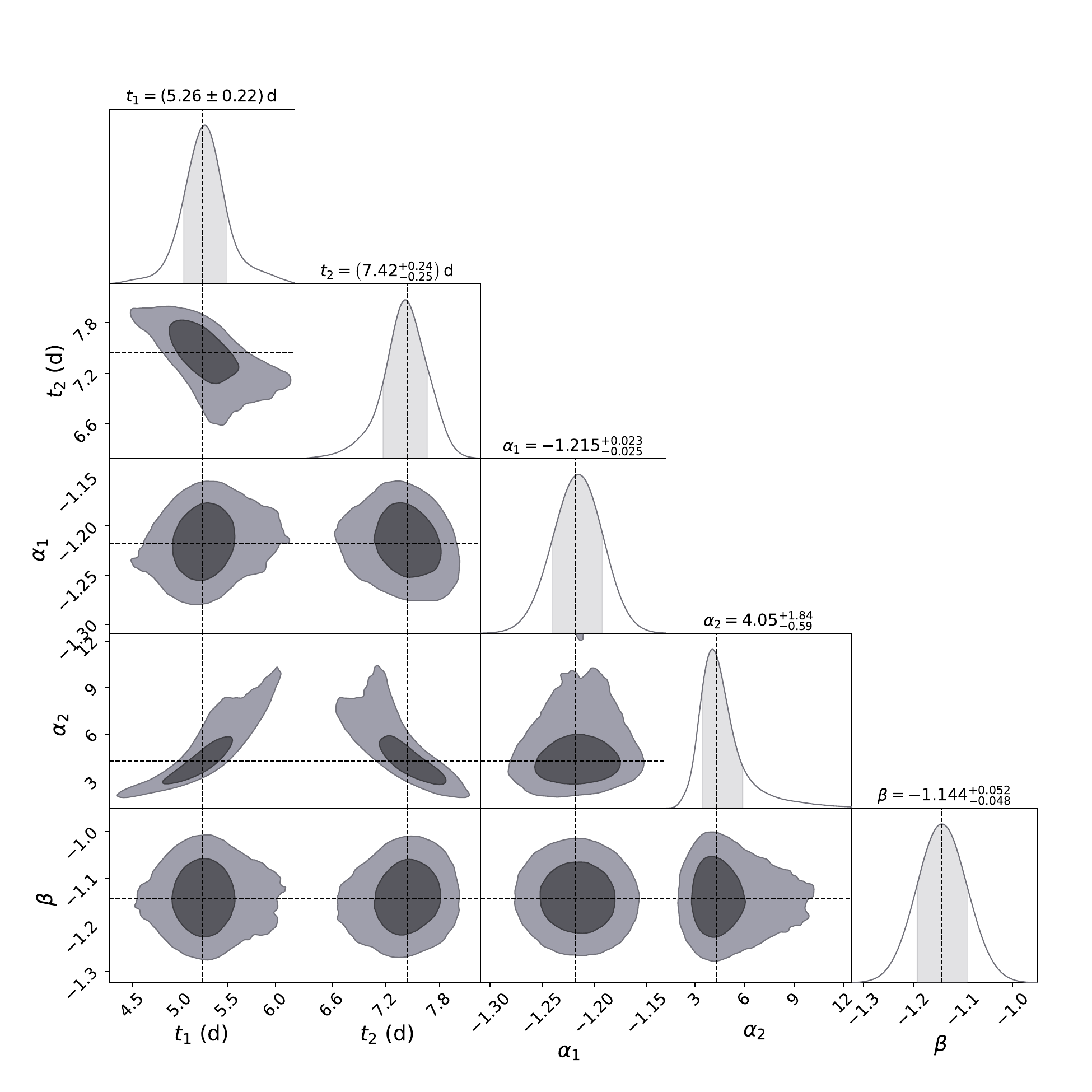}
    \caption{
    Corner plot for the broken powerlaw temporal and spectral fit $F_\nu$\,$\propto$\,$t^\alpha\nu^\beta$ using Equation \ref{eqn:lc} with $\alpha_3 = \alpha_1$ to the multi-band $rzJ$ lightcurves of EP241021a (see \S\ref{sec:temporal}).
    The dashed lines indicate the best fit values in the full chain. The fit has $\chi^2 / \mathrm{dof} = 201 / 76 = 2.58$.
    }
    \label{fig:power-laws-corner-same-slopes}
\end{figure}

The corner plot for both fits are shown separately in Figures \ref{fig:power-laws-corner} and \ref{fig:power-laws-corner-same-slopes}. 
The summary statistics shown in the corner plots are computed by using the maximum of the marginalized posterior as best fit value (presented as the mean value) and the 16th and 84th quantiles are used for the $1\sigma$ uncertainty. This corresponds to the \texttt{MAX\_CENTRAL} option in \texttt{ChainConsumer} \citep{2016JOSS....1...45H}. We note that only parameter where the 50th percentile is not the maximum of the marginalized posterior is the rising slope $\alpha_2$, which has an extremely skewed, non-Gaussian posterior.

\begin{table}
    \centering
    \begin{tabular}{ccc}
    \toprule
        Parameter & $\alpha_1 \neq \alpha_3$ &  $\alpha_1 = \alpha_3$\\
    \midrule
        $\log A$ & $\mathcal{U}(-10, 10)$ & $\mathcal{U}(-10, 10)$\\ 
        $t_1$ & $\mathcal{U}(4, 7)$ & $\mathcal{U}(4, 7)$\\ 
        $t_2$ & $\mathcal{U}(5, 11), t_1 < t_2$ & $\mathcal{U}(5, 11), t_1 < t_2$\\ 
        $\alpha_1$ & $\mathcal{U}(-4, 0)$ & $\mathcal{U}(-4, 0)$\\ 
        $\alpha_2$ & $\mathcal{U}(0, 12.5)$ & $\mathcal{U}(0, 12.5)$\\ 
        $\alpha_3$ & $\mathcal{U}(-4, 0)$ & $-$\\ 
        $\log \Delta_1$ & $\mathcal{U}(-3, -1)$ & $\mathcal{U}(-3, -1.5)$\\ 
        $\log \Delta_2$ & $\mathcal{U}(-3, -1)$ & $\mathcal{U}(-3, -1)$\\ 
        $\beta$ & $\mathcal{U}(-5, 0)$ & $\mathcal{U}(-5, 0)$\\ 
    \bottomrule     
    \end{tabular}
    \caption{Priors for the fit of Equation~\ref{eqn:lc} in Figure~\ref{fig:optlcmag} ($\alpha_1$\,$\neq$\,$\alpha_3$) and Figure~\ref{fig:optlcmag-same-slopes} ($\alpha_1$\,$=$\,$\alpha_3$).}
    \label{tab:priors-light-curve}
\end{table}

\section{Supernova Model Fitting}
\label{appendix:mosfit}

We applied five different \texttt{MOSFiT} models to EP241021a: \textit{i}) \default{} \citep[corresponding to the radioactive decay of $^{56}$Ni;][]{Nadyozhin1994}, \textit{ii}) \csm{} \citep{Chatzopoulos2013, Villar2017}, \textit{iii}) \csmni, \textit{iv}) \magnetar{} \citep{Nicholl2017mosfit}, and \textit{v}) \magni{}. To fit the second component (starting from $\mathrm{MJD}=60609.81$) of EP241021 using \texttt{MOSFiT}, we fix the explosion time to the EP trigger, such that $t_\mathrm{exp}$\,$=$\,$-5.60\,\mathrm{days}$ prior to the onset of the rebrightening. 
The $\gamma$-ray opacity ($\kappa_{\gamma}$) was fixed to $0.027$ cm$^2$ g$^{-1}$ \citep{Cappellaro1997}.
We perform the fit using the dynamic nested sampling approach implemented in \mosfit{} using the \dynesty{} package \citep{Speagle2020, Koposov2023}.
For the \magnetar{} model, we set similar priors as \cite{Gomez2022} to model superluminous supernovae, 
but decreased the range of the magnetar spin period to $P_\mathrm{spin}\in[0.7,10]\,\mathrm{ms}$.
Instead of log-flat priors for $f_\mathrm{Ni}$, $v_\mathrm{ej}$, and $B_\perp$ we chose a flat prior. 
For \csm{} and \csmni{}, we apply similar priors as in \cite{Nicholl2020, Suzuki2021, Chen2023}. We find that none of the models are capable of matching the observed lightcurve. 
For a more detailed overview of the priors for each model see \ref{table:mosfit_priors}.
Relevant posteriors are listed in \ref{table:mosfit_posteriors}.


\input{tables/all_parameters.tex}

\input{tables/all_quantiles.tex}

\end{appendix}
\twocolumn

\end{document}

%% file: tables/all_parameters.tex
\begin{table*}
\caption{Model Parameters (Footnotes: a: $\kappa_{\gamma}$ was not specified in \csm; b: to get this range of $E_{\mathrm{kin}}$, we set $M_\mathrm{ej}\in[0.1,200]\,\mathrm{M}_\odot$ and $v_\mathrm{ej}\in\log([10^3,10^5])\,\mathrm{km}\,\mathrm{s}^{-1}$; c: $\kappa$ was not specified in \csm; d: the \magnetar{} model does not use $f_{\mathrm{Ni}}$; e: the \csm{} model does not use $f_{\mathrm{Ni}}$; f: $v_{\mathrm{ej}}$ in \csmni{} is already embedded in $E_{\mathrm{kin}}$).}
\label{table:mosfit_priors}
\centering
\begin{tabular}{c c c c }
\hline\hline
Prior & \default & \magni & \csmni \\
\hline
$\kappa_{\gamma}\,[\mathrm{cm}^2\,\mathrm{g}^{-1}]$ & $2.70 \times 10^{-2}$ & $2.70 \times 10^{-2}$ & ${2.70 \times 10^{-2}}^{a}$ \\[1mm]
$E_{\mathrm{kin}}\,[10^{51}\,\mathrm{erg}]$ & $-$ & $-$ & $\log([10^{-1}, 10^{2}])^{b}$ \\[1mm]
$s$ & $-$ & $-$ & $[0.00, 2.00]$ \\[1mm]
$\rho\,[\mathrm{g}\,\mathrm{cm}^3]$ & $-$ & $-$ & $\log([10^{-15}, 10^{-11}])$ \\[1mm]
$T\,[\mathrm{K}]$ & $\log([3.00 \times 10^{3}, 10^{4}])$ & $\log([3.00 \times 10^{3}, 10^{4}])$ & $\log([10^{3}, 10^{5}])$ \\[1mm]
$M_{\mathrm{ej}}\,[\mathrm{M}_\odot]$ & $[10^{-1}, 10^{2}]$ & $[10^{-1}, 10^{2}]$ & $[10^{-1}, 2.00 \times 10^{2}]$ \\[1mm]
$\kappa\,[\mathrm{cm}^2\,\mathrm{g}^{-1}]$ & $[5.00 \times 10^{-2}, 3.40 \times 10^{-1}]$ & $[5.00 \times 10^{-2}, 3.40 \times 10^{-1}]$ & $[5.00 \times 10^{-2}, 3.40 \times 10^{-1}]^{c}$ \\[1mm]
$\sigma$ & $\log([10^{-3}, 10^{2}])$ & $\log([10^{-3}, 10^{2}])$ & $\log([10^{-5}, 10^{1}])$ \\[1mm]
$M_{\mathrm{CSM}}\,[\mathrm{M}_\odot]$ & $-$ & $-$ & $[10^{-1}, 2.00 \times 10^{2}]$ \\[1mm]
$f_{\mathrm{Ni}}$ & $[10^{-2}, 5.00 \times 10^{-1}]$ & $[10^{-2}, 5.00 \times 10^{-1}]^{d}$ & $[10^{-2}, 5.00 \times 10^{-1}]^{e}$ \\[1mm]
$r_0\,[\mathrm{AU}]$ & $-$ & $-$ & $\log([10^{-2}, 10^{2}])$ \\[1mm]
$M_{\mathrm{NS}}\,[\mathrm{M}_\odot]$ & $-$ & $[1.50, 1.90]$ & $-$ \\[1mm]
$v_{\mathrm{ej}}\,[\mathrm{km}\,\mathrm{s}^{-1}]$ & $[10^{3}, 10^{5}]$ & $[10^{3}, 10^{5}]$ & $-^{f}$ \\[1mm]
$B_{\mathrm{field}}\,[10^{14}\,\mathrm{G}]$ & $-$ & $[10^{-1}, 1.50 \times 10^{1}]$ & $-$ \\[1mm]
$\delta$ & $-$ & $-$ & $0.00$ \\[1mm]
$P_{\mathrm{spin}}\,[\mathrm{ms}]$ & $-$ & $[7.00 \times 10^{-1}, 10^{1}]$ & $-$ \\[1mm]
$\theta_{\mathrm{PB}}\,[\mathrm{radians}]$ & $-$ & $[0.00, 1.57]$ & $-$ \\
$n$ & $-$ & $-$ & $[7.00, 1.20 \times 10^{1}]$ \\[1mm]
\hline
\end{tabular}
\end{table*}

%% file: tables/all_quantiles.tex
\begin{table*}
\caption{Posteriors of the models \default, \magnetar, \magni, \csm, and \csmni.}
\label{table:mosfit_posteriors}
\centering
\begin{tabular}{lccccc}
\hline\hline
Parameter & \default & \magnetar & \magni & \csm & \csmni \\
\hline 
$P_{\mathrm{spin}}\,[\mathrm{ms}]$ & - & $0.95_{-0.19}^{+0.36}$ & $1.2_{-0.4}^{+0.8}$ & - & - \\[1mm]
$v_{\mathrm{ej}}\,[\mathrm{km}\,\mathrm{s}^{-1}]$ & $97000_{-4400}^{+2200}$ & $95800_{-5500}^{+3000}$ & $96900_{-4700}^{+2300}$ & $79400_{-5300}^{+5700}$ & - \\[1mm]
$\log(E_{\mathrm{kin}}\,[10^{51}\,\mathrm{erg}])$ & - & - & - & - & $1.96_{-0.06}^{+0.03}$ \\[1mm]
$\log(T\,[\mathrm{K}])$ & $3.88_{-0.04}^{+0.05}$ & $3.78_{-0.04}^{+0.05}$ & $3.85_{-0.04}^{+0.04}$ & $4.82_{-0.04}^{+0.04}$ & $4.13_{-0.03}^{+0.03}$ \\[1mm]
$M_{\mathrm{NS}}\,[\mathrm{M}_\odot]$ & - & $1.76_{-0.14}^{+0.10}$ & $1.70_{-0.13}^{+0.13}$ & - & - \\[1mm]
$M_{\mathrm{ej}}\,[\mathrm{M}_\odot]$ & $9.9_{-0.9}^{+1.0}$ & $36.8_{-7.2}^{+14.7}$ & $12.4_{-2.1}^{4.3}$ & $118_{-60}^{+54}$ & $2.9_{-0.5}^{+0.6}$ \\[1mm]
$B_{\mathrm{field}}\,[10^{14}\,\mathrm{G}]$ & - & $7.5_{-1.7}^{+2.9}$ & $11.9_{-2.2}^{+2.00}$ & - & - \\[1mm]
$M_{\mathrm{CSM}}\,[\mathrm{M}_\odot]$ & - & - & - & $2.1_{-0.5}^{+0.4}$ & $95_{-65}^{+66}$ \\[1mm]
$\log(\rho\,[\mathrm{g}\,\mathrm{cm}^3])$ & - & - & - & $-12.0_{-0.8}^{+0.6}$ & $-12.8_{-0.8}^{+1.0}$ \\[1mm]
$r_0\,[\mathrm{AU}]$ & - & - & - & $0.8_{-0.4}^{+0.6}$ & $1.2_{-0.5}^{+0.5}$ \\[1mm]
$f_{\mathrm{Ni}}$ & $0.47_{-0.03}^{+0.02}$ & - & $0.32_{-0.14}^{+0.12}$ & - & $0.33_{-0.17}^{+0.12}$ \\[1mm]
$A_{V,z}\,[\mathrm{mag}]$ & $0.00_{-0.00}^{0.01}$ & $0.00_{-0.00}^{0.01}$ & $0.00_{-0.00}^{0.01}$ & $2.49_{-0.08}^{0.08}$ & $1.68_{-0.12}^{0.11}$ \\[1mm]
\hline
\end{tabular}
\end{table*}